\documentclass[longauth]{aa}  

\usepackage{graphicx}
\usepackage{booktabs}
\usepackage{lscape} 
\usepackage{txfonts}
\usepackage{multirow}
\usepackage{xcolor}
\usepackage{siunitx}
\usepackage{textcomp}
\usepackage{tabularx}
\usepackage{subcaption}
\usepackage{afterpage}
\usepackage{sidecap}
\usepackage{adjustbox}      
\begin{document} 
\newcommand\pp[1]{%
\renewcommand\arraystretch{1}%
\begin{tabular}{@{}l@{}}#1\end{tabular}}

   \title{HYPERION: broad-band X-ray-to-near-infrared emission of Quasars in the first billion years of the Universe}
   \titlerunning{Broad-band emission of high-z QSOs}

   \author{I. Saccheo \inst{1,2}
          \and A. Bongiorno \inst{2}
          \and E. Piconcelli \inst{2}
          \and L. Zappacosta \inst{2}
          \and M. Bischetti \inst{3,4}
          \and V. D'Odorico \inst{3,5,6}
          \and C. Done \inst{7}
          \and M. J. Temple \inst{7}
          \and V. Testa \inst{2}
          \and A. Tortosa \inst{2}
          \and M. Brusa \inst{8, 9}
          \and S. Carniani \inst{5}
          \and F. Civano \inst{10}
          \and A. Comastri \inst{9}
          \and S. Cristiani \inst{3,6,11}
          \and D. De Cicco \inst{12, 13, 14}
          \and M. Elvis \inst{15}
          \and X. Fan \inst{16}
          \and C. Feruglio \inst{3,6}
          \and F. Fiore \inst{3,6}
          \and S. Gallerani \inst{5}
          \and E. Giallongo \inst{2}
          \and R. Gilli \inst{9}
          \and A. Grazian \inst{17}
          \and M. Guainazzi \inst{18}
          \and F. Haardt \inst{19,20,21}
          \and R. Maiolino \inst{22,23,24}
          \and N. Menci \inst{2}
          \and G. Miniutti \inst{25}
          \and F. Nicastro \inst{2}
          \and M. Paolillo \inst{12,14}
          \and S. Puccetti \inst{26}
          \and F. Salvestrini \inst{3}
          \and R. Schneider \inst{2,27,28,29}
          \and F. Tombesi \inst{30,31,2}
          \and R. Tripodi \inst{32,3,4,6}
          \and R. Valiante \inst{2,28}
          \and L. Vallini \inst{9}
          \and E. Vanzella \inst{9}
          \and G. Vietri \inst{33}
          \and C. Vignali \inst{8,9}
          \and F. Vito \inst{9}
          \and M. Volonteri \inst{34}
          \and F. La Franca \inst{1,2}
          }
   \institute{
   Dipartimento di Matematica e Fisica, Università Roma Tre, Via della Vasca Navale 84, 00146 Roma, Italy.
   \and INAF - Osservatorio astronomico di Roma, Via Frascati 33, I-00040 Monte Porzio Catone, Italy.
   \and INAF - Osservatorio Astronomico di Trieste, Via G. Tiepolo 11, I-34143 Trieste, Italy.
   \and Dipartimento di Fisica, Sezione di Astronomia, Università di Trieste, via Tiepolo 11, I-34143 Trieste, Italy.
   \and Scuola Normale Superiore, Piazza dei Cavalieri 7, I-56126 Pisa, Italy.
   \and IFPU - Institute for Fundamental Physics of the Universe, via Beirut 2, I-34151 Trieste, Italy.
   \and Centre for Extragalactic Astronomy, Department of Physics, Durham University, South Road, Durham DH1 3LE, UK.
   \and Dipartimento di Fisica e Astronomia ‘Augusto Righi’, Università degli Studi di Bologna, via P. Gobetti, 93/2, 40129 Bologna, Italy.
   \and INAF - Osservatorio di Astrofisica e Scienza dello Spazio di Bologna, via Piero Gobetti, 93/3, I-40129 Bologna, Italy.
   \and NASA Goddard Space Flight Center, Greenbelt, MD 20771, USA.
   \and INFN - National Institute for Nuclear Physics, via Valerio 2, I-34127 Trieste, Italy.
   \and Department of Physics, University of Napoli ‘Federico II’, via Cinthia 9, 80126 Napoli, Italy.
  \and Millennium Institute of Astrophysics (MAS), Nuncio Monse\~nor Sotero Sanz 100, Providencia, Santiago, Chile. 
  \and INAF - Osservatorio Astronomico di Capodimonte, via Moiariello 16, 80131 Napoli, Italy 
  \and Center for Astrophysics - Harvard \& Smithsonian, Cambridge, MA 02138, USA.
  \and Steward Observatory, University of Arizona, Tucson, Arizona, USA.
 \and INAF - Osservatorio Astronomico di Padova, Vicolo dell'Osservatorio 5, I-35122, Padova, Italy. 
 \and European Space Agency, ESTEC, Keplerlaan 1, 2201 AZ Noordwijk, The Netherlands
 \and DiSAT, Università degli Studi dell’Insubria, Via Valleggio 11, I-22100 Como, Italy.
 \and INFN - Sezione di Milano-Bicocca, Piazza della Scienza 3, I-20126 Milano, Italy. 
\and INAF - Osservatorio Astronomico di Brera, Via E. Bianchi 46, I-23807 Merate, Italy.
\and Cavendish Laboratory, University of Cambridge, 19 J. J. Thomson Ave., Cambridge CB3 0HE, UK.
\and Kavli Institute for Cosmology, University of Cambridge, Madingley Road, Cambridge CB3 0HA, UK.
\and Department of Physics \& Astronomy, University College London, Gower Street, London WC1E 6BT, UK.
\and Centro de Astrobiología (CAB), CSIC-INTA, Camino Bajo del Castillo s/n, ESAC campus, 28692 Villanueva de la Cañada, Spain.
\and ASI - Agenzia Spaziale Italiana, Via del Politecnico snc, I-00133 Roma, Italy.
\and Dipartimento di Fisica, Università di Roma La Sapienza, Piazzale Aldo Moro 2, I-00185 Roma, Italy.
\and INFN - Sezione Roma1, Dipartimento di Fisica, Università di Roma La Sapienza, Piazzale Aldo Moro 2, I-00185 Roma, Italy.
\and Sapienza School for Advanced Studies, Viale Regina Elena 291, I- 00161 Roma, Italy. 
\and Physics Department, Tor Vergata University of Rome, Via della Ricerca Scientifica 1, 00133 Rome, Italy.
\and INFN - Rome Tor Vergata, Via della Ricerca Scientifica 1, 00133 Rome, Italy.
\and Department of Astronomy, University of Maryland, College Park, MD 20742, USA.
\and University of Ljubljana, Department of Mathematics and Physics, Jadranska ulica 19, SI-1000 Ljubljana, Slovenia.
\and INAF - Istituto di Astrofisica Spaziale e Fisica Cosmica Milano, Via A. Corti 12, 20133 Milano, Italy.
\and Institut d’Astrophysique de Paris, Sorbonne Université, CNRS, UMR 7095, 98 bis bd Arago, 75014 Paris, France.
}

   \date{Received 25/05/2024; accepted 18/10/2024}
  \abstract
   {}
   {We aim at characterizing the X-ray-to-optical/near-infrared broad-band emission of luminous QSOs in the first Gyr of cosmic evolution to understand whether they exhibit differences  compared to the lower-\textit{z} QSO population. Our goal is also to provide for these objects a reliable and uniform catalog of SED fitting derivable properties such as bolometric and monochromatic luminosities, Eddington ratios, dust extinction, strength of the hot dust emission.}
   {
   We gather all available photometry, covering from XMM-Newton proprietary data in the X-ray to rest-frame NIR wavelengths for the 18 QSOs in the HYPERION samples ($6.0 \leq z \leq 7.5$). For sources lacking a uniform NIR coverage, we conduct NIR observations in the J, H, and K bands. To increase the statistical robustness of our analysis across the UV-to-NIR region, we add to our sample 36 additional sources from the E-XQR-30 sample with 5.7 $\lesssim z \lesssim$ 6.6.

   We characterize the X-ray/UV emission of each QSO using average SEDs from luminous Type 1 sources and calculate bolometric and monochromatic luminosities.
   Finally we construct a mean SED extending from the X-rays to the NIR bands.} 
   {We find that the UV-optical emission of these QSOs can be modelled with templates of $z\sim$2 luminous QSOs. We observe that the bolometric luminosities derived adopting some bolometric corrections at 3000 \AA\ ($BC_{3000\text{\AA}}$) largely used in the literature are slightly overestimated by 0.13 dex as they also include reprocessed IR emission. We estimate a revised value, i.e.  $BC_{3000\text{\AA}}=3.3 $ which can be used for deriving $L_\text{bol}$ in \textit{z} $\geq$ 6 QSOs. A sub-sample of 11 QSOs is provided with rest-frame NIR photometry, showing a broad range of hot dust emission strength, with two sources exhibiting low levels of emission. Despite potential observational biases arising from non-uniform photometric coverage and selection biases, we produce a X-ray-to-NIR mean SED for QSOs at \textit{z} $\gtrsim$ 6, revealing a good match with templates of lower-redshift, luminous QSOs up to the UV-optical range, with a slightly enhanced contribution from hot dust in the NIR.}
   {}


   \maketitle
%

\section{Introduction}
Thanks to the use of wide field optical surveys such as SDSS, the Canada-France High-z Quasar Survey, (CFHQS), or Pan-STARRs, to date about 300 quasars (QSOs) have been discovered at $z\sim 6-7.6$ \citep[e.g.][and references therein]{Jiang:2016,Banados:2016, Mazzucchelli:2017, Fan:2023}. These QSOs are powered by accretion into supermassive black holes (SMBHs) with masses $M_\text{SMBH} > 10^{8}-10^{9}M_{\odot}$ and exhibit bolometric luminosities $L_\text{bol} > 10^{46}\,erg\,s^{-1}$  \citep[e.g.][]{Mazzucchelli:2017, Shen:2019, Mazzucchelli:2023}. \\
The presence of fully grown SMBHs already at the Epoch of Reionization (EoR, $ z \gtrsim 5.5$) presents a challenge to theoretical models aiming to understand their formation in the early stages of cosmic evolution \citep[e.g][]{Volonteri:2010, Inayoshi:2020, Volonteri:2021}. 
To grow SMBHs at such high masses in a short amount of time, models usually employ two distinct pathways: a) super-Eddington accretion \citep[e.g.][]{Madau:2014, Volonteri:2015, Pezzulli:2016}; b) massive BH seeds with $M_{\text{BH, seed}} \sim 10^{3}-10^{4} M_{\odot}$ \citep[e.g.][]{Volonteri:2010, Valiante:2016}. The a) scenario allows SMBH formation from lower-mass seeds, with $M_{\text{BH, seed}} \sim 10^{2} M_{\odot}$, due to PopIII stars remnants. Scenario b) instead allows an Eddington-limited accretion regime.

The upcoming Euclid and LSST surveys,  which, according to the latest luminosity functions \citep[e.g.][]{Shen:2020} are expected to uncover thousands of new QSOs at $z\geq 6$, will provide a complete census of these primeval sources and may help to constrain their  evolutionary history.\\ 
So far, one finding of these first QSOs is that their UV-to-Near Infrared (NIR) broad-band emission resembles the one observed in lower redshift counterparts \citep[e.g.][]{Fan:2004, Iwamuro:2004, Reed:2019}; moreover, by analyzing the UV-to-optical spectra of 50 QSOs at $z \geq 5.7$ \cite{Shen:2019} found no significant major difference with the ones observed in the SDSS sample \citep[][]{Vanden-Berk:2001}. Nevertheless, some distinctions are reported and include faster disk winds traced by CIV blueshifts \citep[][]{Meyer:2019, Schindler:2020, Yang:2021} and a higher occurrence of Broad Absorption lines (BAL) \citep[][]{Bischetti:2022} in high-\textit{z} QSOs compared to lower redshift objects of similar luminosities. 
At higher energies, the monochromatic UV to X-ray luminosity ratio shows no significant redshift evolution \citep[][]{Vito:2019, Wang:2021}, following a trend akin to lower-\textit{z} QSOs \citep[e.g.][]{Vignali:2003, Just:2007, Lusso:2010}, although \cite{Zappacosta:2023} report evidence for a mild deviation which needs to be further investigated.
In contrast, in the X-ray band, \cite{Zappacosta:2023} recently presented the results of the X-ray spectral analysis conducted on a sample of \textit{z} $\geq$ 6 QSOs, namely the HYPERION QSOs, showing that these highly accreting sources have a significant steeper photon index $\Gamma$ than analogously luminous \textit{z} $<$ 6 QSOs \citep[see also][]{Vito:2018}. This finding hints at a redshift evolution of $\Gamma$, possibly linked to a different accretion process taking place in the region closest to the SMBH.\\
To date, $L_\text{bol}$ of high-redshift QSOs have been mainly computed through optical bolometric corrections, \citep[e.g][]{Reed:2019, Yang:2021, Mazzucchelli:2023}, that have been shown to be roughly constant over a wide range of luminosities and redshifts \citep[][but see also \cite{Trakhtenbrot:2012}]{Duras:2020}. However bolometric corrections have an intrinsic dispersion of $\sim 0.25$ dex \citep[][]{Duras:2020}, that can be accounted for the scatter in the slopes of the power laws describing individual QSOs optical continua and thus they do not necessarily provide the most reliable results on individual sources. Moreover, most measurements of the optical luminosities do not account for dust extinction which could lead to their underestimation. \\
In this work we systematically investigated the X-ray-to-NIR continuum emission of a sample of the 18 HYPERION QSOs, which benefit from exhaustive multi-wavelength coverage between these bands. We also included 36 additional sources with comparable redshift and luminosity distributions with rest-frame UV-to-NIR photometry to increase the statistical significance of our analysis. Our main goal is to provide accurate  measurements of accretion disk related properties (i.e. bolometric and monochromatic luminosities, Eddington ratios) via Spectral Energy Distribution \citep[SED, e.g.][]{Elvis:1994} fitting and thus to construct a homogeneous reference catalog. In addition, we derive a mean SED of these sources ranging from the X-ray to the NIR and compare it to that measured for low redshift QSOs \citep[e.g.][]{Vanden-Berk:2001, Krawczyk:2013}. 
Throughout this paper we use a standard Flat $\Lambda$CDM cosmology with $H_{0}= 70$ Km/s Mpc$^{-1}$ and $\Omega_{0} = 0.27$. Unless otherwise stated, uncertainties are reported at 68\% confidence level and upper limits in the photometry are reported at the 3$\sigma$ level.

\section{The sample}

\begin{table}
\centering
\addtolength{\tabcolsep}{-3.2pt}  
\scalebox{0.90}{
\begin{tabular}{lccccc}
Name & Redshift & Log($M_{SMBH}$) & Log($L_{BC}$) &
$M_{\rm s, Edd}$ & Log($F_{0.3-7keV}$)\\ 
& & [$M_{\odot}$] & [erg/s] & [$M_{\odot}$] & [erg/s cm$^{-2}$]\\
\toprule
J1342+0928 & 7.541  &   8.90  & 47.19  & 19120   & -15.52 $\pm$ 0.06  \\    
J1007+2115 & 7.514  &   9.18  & 47.30  & 32460   & -15.16 $\pm$ 0.07  \\               
J1120+0641 & 7.087  &   9.41  & 47.30  & 18230   & -15.16 $\pm$ 0.07  \\              
J0038-1527 & 7.021  &   9.14  & 47.36  &  7983   &  $<$-16.09         \\              
J0252-0503 & 7.0    &   9.15  & 47.12  &  7679   & -14.12 $\pm$ 0.05  \\                  
J0020-3653 & 6.834  &   9.24  & 47.16  &  5753   & -15.33 $\pm$ 0.08  \\                 
J0411-0907 & 6.824  &   8.80  & 47.31  &  2019   & -14.49 $\pm$ 0.09  \\                 
J0244-5008 & 6.724  &   9.08  & 47.19  &  2814   & -15.33 $\pm$ 0.03  \\                    
J231-20.8  & 6.587  &   9.50  & 47.31  &  4708   & -15.32 $\pm$ 0.05  \\                
J036+03.0  & 6.533  &   9.49  & 47.33  &  3776   & -15.63 $\pm$ 0.09  \\                    
J0224-4711 & 6.526  &   9.36  & 47.53  &  2730   & -14.23 $\pm$0.07   \\                     
J011+09    & 6.444  &   9.15  & 47.12  &  1279   & -15.44  $\pm$ 0.20 \\                  
J1148+5251 & 6.422  &   9.74  & 47.57  &  4627   & -14.76 $\pm$ 0.35  \\                 
J083+11.8  & 6.346  &   9.32  & 47.16  &  1324   & -14.53 $\pm$ 0.06  \\                 
J0100+2802 & 6.3    &  10.04  & 48.24  &  5799   & -14.25 $\pm$ 0.05  \\   
J025-33    & 6.294  &   9.57  & 47.39  &  1392   & -15.62 $\pm$ 0.06  \\    
J0050+3445 & 6.245  &   9.68  & 47.29  &  2072   & -14.50 $\pm$ 0.05  \\               
J029-36    & 6.02   &   9.82  & 47.39  &  1220   & -15.44 $\pm$ 0.05  \\   
           
\bottomrule

\end{tabular}
}
\caption{The HYPERION sample. Redshift and $M_{SMBH}$ reported have been all estimated from the MgII line, see \cite{Zappacosta:2023} for further references. X-ray fluxes from \cite{Zappacosta:2023} and \cite{Tortosa:2024b}.}
\label{table:coordinates}
\end{table}

To perform a reliable X-ray-to-NIR description of QSOs at z$>$6 we considered the sources in the HYPerluminous quasars at the Epoch of ReionizatION \citep[HYPERION;][] {Zappacosta:2023} sample, which have been recently targeted by a $\sim$ 700 hours XMM-Newton Heritage Programme (PI L. Zappacosta) providing them the best quality X-ray data to date.
The HYPERION sample includes the sources whose SMBHs experienced the fastest mass growth rates during their formation. In particular, assuming an exponential 
continuous growth at the Eddington rate limit, the HYPERION QSOs have been selected as the
luminous ($L_\text{bol}>10^{47}$ erg/s) \textit{z}$>$6 sources whose SMBH required, to assemble, a seed BH mass (hereafter $M_{\rm s, Edd}$) of $>$1000 $M_{\odot}$ \citep[formed at z=20,][] {Valiante:2016}. $M_{\rm s, Edd}$ is a proxy of the growth rate history experienced by SMBHs; more specifically the HYPERION selection requires that 
\begin{equation*}
    M_{\rm s, Edd}  = M_\text{SMBH}\times \exp(-t/t_{s}) \geq 1000M_{\odot}
\end{equation*}
where $t$ is the time elapsed since the seed formation and $t_{s} = 0.45 \varepsilon(1-\varepsilon)^{-1}\lambda_\text{Edd}^{-1}f_\text{duty}^{-1}$ is the e-folding time, where $\lambda_\text{Edd} = \frac{L_\text{bol}}{L_\text{Edd}}$ (=1),  $f_\text{duty}$(=1) and $\varepsilon$ are the Eddington ratio, the duty cycle,  i.e. the fraction of time during which  the QSO is active, and the radiative efficiency, i.e. the fraction of accreting mass radiated away, respectively. 

In Tab. \ref{table:coordinates} we present the HYPERION sources together with some physical parameters i.e. \textit{z}, $M_\text{SMBH}$, $L_\text{BC}$, i.e. bolometric luminosities derived assuming a 3000 \AA\ bolometric correction, $M_{\rm s, Edd}$ and measured X-ray fluxes \citep[see][]{Zappacosta:2023, Tortosa:2024b}.\\
HYPERION aims at the highest quality, and hence most reliable, determination of the X-ray nuclear properties of QSOs at EoR. As mentioned, results from the analysis of first-year observations, reported in \cite{Zappacosta:2023}, indicates an average photon index $\Gamma \approx 2.4 \pm 0.1$ which is significantly steeper than the $\Gamma \sim 1.8-2 $ typically observed in AGN at lower redshifts \citep[e.g.][]{Vignali:2005, Piconcelli:2005, Dadina:2008,Zappacosta:2018, Zappacosta:2020}. 
Moreover the Far Infrared and sub-millimetre region of the SED of these objects as well as dust and gas properties in the host of HYPERION sources are being investigated through ALMA and NOEMA data in a series of papers \citep[][]{Tripodi:2023, Tripodi:2023b, Feruglio:2023}.\\
To increase the statistical significance of our analysis in bands UV-to-NIR we included as a complementary sample the Ultimate XSHOOTER legacy survey of quasars at z $\sim$ 5.8 - 6.6 \citep[][hereafter E-XQR-30]{DOdorico:2023}, which consists of 42 bright sources ($i_{mag}<20$) at $z\gtrsim 5.7$. Notably, six QSOs, J231-20.8, J0224-4711, J029-36, J036+03.0, J025-33 and J0100+2802, belong to both HYPERION and E-XQR-30, and thus the considered E-XQR-30 sub-sample consists of 36 sources.

\subsection{Multiwavelength photometric data}
In addition to the good quality X-ray fluxes measured 
for all HYPERION QSOs by \cite{Zappacosta:2023} and \cite{Tortosa:2024b}, to construct the rest-frame X-ray-to-NIR SED of the HYPERION QSOs, we gathered all Near and Mid IR photometric data available in the literature, specifically from the z, Y, J, H, K filters plus the four WISE (3.4, 4.6, 12 and 22 \textmu m, called W1, W2, W3 and W4 respectively) bands. Filters bluer than the z one correspond to rest-frame wavelengths shorter than the Ly$\alpha$ for all sources and therefore were not considered.\\ 
Only 2 sources, J1148+5251 and J0100+2802, have complete photometric coverage from z to W4, while for 12 QSOs we found photometry in all bands from z to W2; 3 sources lacked of data in the H and K filters, and for the remaining 4 QSOs only the H band was missing.
For roughly half of the sources in the sample, multiple data are available in the same band for at least one filter. When possible we preferred data obtained from targeted observations or data already published over data from surveys.  For what concerns survey catalogs, we made use of the \textit{Vista Hemisphere Survey} \citep[VHS,][]{McMahon:2021}, UKIDSS-LAS \citep[][]{Lawrence:2012}, the \textit{Dark Energy Survey} \citep[DES,][we used the 3 arcsec aperture magnitudes]{Abbott:2021} and the Pan-STARRS survey \citep[][PSF magnitudes reported]{Kaiser:2002}. Two sources (J0411-0907 and J036+03.0) have their Y band covered by multiple catalogs, with their values in agreement within 0.1 magnitudes. In this case we decided to use the DES values.\\

For the WISE data, we relied on the unWISE Catalog by \cite{Schlafly:2019} for W1 and W2 filters, and on the AllWISE catalog \citep[][]{Cutri:2013} for W3 and W4. Given the wide PSF in these bands, ($\sim$ 7.3 and 12 arcsec respectively), and the possibility of contamination from nearby projected  companions, we did not use them in the fitting routine.\\
J1120+06 and J1148+52 have Spitzer IRAC and MIPS observations in 3.6 and  4.5 \textmu m filters from \cite{Jiang:2006} and  3.6, 4.5, 5.8, 8.0 and 24 \textmu m bands from \cite{Barnett:2015} respectively. For these two QSOs, as well as for the 6 E-XQR-30 sources with available Spitzer data, we used the 3.6 and 4.5 \textmu m bands instead of W1 and W2.
Moreover, for J1120+0641 we also included the 1 \textmu m luminosity derived by \cite{Bosman:2023} from its JWST/MIRI spectrum, i.e. $\lambda L_{1\mu m} = 1.37 \pm 0.08 \times 10^{46}$ erg/s.\\
The above described multi-wavelength photometric data has been complemented by additional proprietary NIR observations (see Appendix \ref{sec:proprietary_observations}) with the goal of achieving homogeneous and complete coverage of the emission from these sources. 
A detailed description of the NIR observations can be found in Appendix \ref{sec:proprietary_observations} while  the used AB magnitudes and their references of the HYPERION sample are reported in Tab. \ref{table:photometry_master}.
The multi-wavelength data of the E-XQR-30 sample are instead reported in Tab. \ref{table:photometry_xqr30}. For more details on their properties we refer to \cite{Bischetti:2022,DOdorico:2023} (J,H and K bands), \cite{Ross:2020} (z, Y and W1 to W4 bands), and \cite{Leipski:2014} (Spitzer IRAC 3.6, 4.5, 5.8 and Spitzer MIPS 8.0 and 24 \textmu m).

\begin{table*}[h]
\begin{tabular}{lcccccccc}
\toprule
              Name &   $L_\text{bol}$ &             $\lambda L_{2500\text{\AA}}$ &            $\lambda L_{3000\text{\AA}}$ &            $\lambda L_{4400\text{\AA}}$ &            $\lambda L_{5100\text{\AA}}$ &  E[B-V] & 
              $\lambda_\text{Edd}$  & $\lambda L_{1keV}$\\
\midrule
J1342+0928	&	47.05 $\pm $ 0.04	&	46.60 $\pm $ 0.02	&	46.46 $\pm $ 0.02	&	46.20 $\pm $ 0.05	&	46.14 $\pm $ 0.07	&	0.01	&	1.11 $\pm $ 0.38	&	44.93 $\pm $ 0.09 \\
J1007+2115	&	47.19 $\pm $ 0.07	&	46.75 $\pm $ 0.03	&	46.67 $\pm $ 0.02	&	46.50 $\pm $ 0.03	&	46.44 $\pm $ 0.04	&	0.03	&	0.81 $\pm $ 0.16	&	44.87 $\pm $ 0.18 \\
J1120+0641	&	47.22 $\pm $ 0.05	&	46.84 $\pm $ 0.05	&	46.76 $\pm $ 0.05	&	46.59 $\pm $ 0.01	&	46.49 $\pm $ 0.01	&	0.05	&	0.50 $\pm $ 0.28	&	45.19 $\pm $ 0.07 \\
J0038-1527	&	47.25 $\pm $ 0.06	&	46.87 $\pm $ 0.03	&	46.78 $\pm $ 0.03	&	46.53 $\pm $ 0.02	&	46.42 $\pm $ 0.03	&	0.03	&	1.00 $\pm $ 0.23	&	44.96 $\pm $ 0.16\\
J0252-0503	&	47.00 $\pm $ 0.06	&	46.59 $\pm $ 0.04	&	46.48 $\pm $ 0.04	&	46.24 $\pm $ 0.04	&	46.15 $\pm $ 0.06	&	0.01	&	0.56 $\pm $ 0.08	&	44.95 $\pm $ 0.19\\
J0020-3653	&	47.10 $\pm $ 0.06	&	46.69 $\pm $ 0.05	&	46.66 $\pm $ 0.04	&	46.46 $\pm $ 0.02	&	46.35 $\pm $ 0.03	&	0.02	&	0.58 $\pm $ 0.13	&	45.21 $\pm $ 0.09\\
J0411-0907	&	47.17 $\pm $ 0.04	&	46.74 $\pm $ 0.02	&	46.66 $\pm $ 0.02	&	46.47 $\pm $ 0.02	&	46.43 $\pm $ 0.03	&	0.01	&	1.84 $\pm $ 0.24	&	44.82 $\pm $ 0.08\\
J0244-5008	&	47.07 $\pm $ 0.04	&	46.58 $\pm $ 0.03	&	46.54 $\pm $ 0.02	&	46.41 $\pm $ 0.01	&	46.37 $\pm $ 0.02	&	0.01	&	0.77 $\pm $ 0.27	&	45.31 $\pm $ 0.05\\
J231.6-20.8	&	47.13 $\pm $ 0.05	&	46.70 $\pm $ 0.04	&	46.59 $\pm $ 0.07	&	46.29 $\pm $ 0.06	&	46.21 $\pm $ 0.10	&	0.01	&	0.34 $\pm $ 0.15	&	44.73 $\pm $ 0.13\\
J036.5+03.0	&	47.23 $\pm $ 0.04	&	46.79 $\pm $ 0.03	&	46.72 $\pm $ 0.03	&	46.50 $\pm $ 0.02	&	46.40 $\pm $ 0.03	&	0.0	&	0.44 $\pm $ 0.13	&	44.61 $\pm $ 0.14 \\
J0224-4711	&	47.47 $\pm $ 0.05	&	47.04 $\pm $ 0.02	&	47.04 $\pm $ 0.02	&	46.82 $\pm $ 0.01	&	46.75 $\pm $ 0.01	&	0.07	&	0.75 $\pm $ 0.19	&	45.35 $\pm $ 0.06\\
J011+09	&	46.87 $\pm $ 0.04	&	46.39 $\pm $ 0.02	&	46.42 $\pm $ 0.02	&	46.06 $\pm $ 0.04	&	45.92 $\pm $ 0.05	&	0.01	&	0.42 $\pm $ 0.04	&	44.86 $\pm $ 0.11 \\
J1148+5251	&	47.48 $\pm $ 0.04	&	46.95 $\pm $ 0.02	&	46.91 $\pm $ 0.02	&	46.81 $\pm $ 0.01	&	46.76 $\pm $ 0.01	&	0.02	&	0.44 $\pm $ 0.05	&	45.30 $\pm $ 0.09 \\
J083.8+11.8	&	47.13 $\pm $ 0.04	&	46.71 $\pm $ 0.03	&	46.64 $\pm $ 0.03	&	46.48 $\pm $ 0.02	&	46.38 $\pm $ 0.03	&	0.01	&	0.51 $\pm $ 0.16	&	44.41 $\pm $ 0.07 \\
J0100+2802	&	47.99 $\pm $ 0.03	&	47.54 $\pm $ 0.03	&	47.57 $\pm $ 0.06	&	47.37 $\pm $ 0.01	&	47.32 $\pm $ 0.01	&	0.0	&	0.70 $\pm $ 0.11	&	45.79 $\pm $ 0.03 \\
J025-33	&	47.38 $\pm $ 0.04	&	46.93 $\pm $ 0.01	&	46.89 $\pm $ 0.01	&	46.69 $\pm $ 0.01	&	46.64 $\pm $ 0.01	&	0.01	&	0.52 $\pm $ 0.04	&	45.06 $\pm $ 0.14 \\
J0050+3445	&	47.24 $\pm $ 0.06	&	46.84 $\pm $ 0.02	&	46.78 $\pm $ 0.03	&	46.55 $\pm $ 0.01	&	46.51 $\pm $ 0.01	&	0.04	&	0.29 $\pm $ 0.06	&	44.76 $\pm $ 0.09 \\
J029-36	&	47.17 $\pm $ 0.06	&	46.66 $\pm $ 0.03	&	46.60 $\pm $ 0.02	&	46.48 $\pm $ 0.01	&	46.42 $\pm $ 0.01	&	0.02	&	0.17 $\pm $ 0.03	&	45.03 $\pm $ 0.09 \\
\bottomrule
\end{tabular}
\caption{Logarithm of bolometric and monochromatic luminosities of the HYPERION sample in units of [erg/s] derived from the lum-K13 best-fit. Monochromatic luminosities are corrected for the E[B-V]. The Eddington ratios $\lambda_\text{Edd}$ have been computed assuming the black hole masses $M_\text{BH}$ reported in Tab. \ref{table:coordinates}.
$L_{1keV}$ was derived from the observed 0.3-7 keV fluxes by adopting the photon index $\Gamma$ reported in \cite{Zappacosta:2023} and \cite{Tortosa:2024b}}
\label{table:luminosities}
\end{table*}

\section{SED fitting}
\label{sec:sed_fitting}
\subsection{UV-to-NIR SED fittitng}
\label{sec:sed_fitting_uv}
To analyze the SEDs for our QSOs, we performed SED fitting  on the UV-to-NIR data using empirically derived pure AGN templates. Indeed, given the high-luminosity regime probed by the sources in our sample, we only accounted for the QSO emission, assuming a negligible contribution on the photometric points by the host-galaxy \citep[e.g.][]{Shen:2011}.\\ 
Among the different templates in the literature, we utilized the two mean SEDs derived from samples which best match the luminosity distribution of our sources. These templates are the mean SED computed by \cite{Krawczyk:2013} for their high luminosity subsample ($Log(L_{2500\text{\AA}}) \geq 45.85$ erg/s, 0.5 $\lesssim z \lesssim$ 4.8), hereafter lum-K13, and the mean SED computed using the WISSH hyper-luminous QSOs \citep[][]{Saccheo:2023}, hereafter WISSH-S23, that span an even closer luminosity range to the analyzed sample ($\log(L_\text{bol}) > 47$ erg/s, 1.8 $\leq z \leq$ 4.8).\\
To account for intrinsic variations in the QSO SEDs that are not captured by the mean SED, we quantified the typical scatter of observed QSO SEDs relative to their average template. This step is needed because, even when restricting to very luminous sources, QSO SEDs show significant variations in their shapes \citep[e.g.][]{Richards:2006, Temple:2021b}. Since the amount of scatter depends on the wavelength, this corresponds to assigning greater weights to the points where the SEDs show fewer variations.
To calculate this scatter, we normalized the luminosity points of the QSOs originally used to compute the lum-K13 and WISSH-S23 templates using their $L_{\text{bol}}$. These normalized points were then binned into equally spaced bins on a logarithmic scale. The scatter in each bin was computed using the Median Absolute Deviation (MAD), i.e. $\sigma_{\text{SED}} = \text{MAD}(\bar{y} - f)$, where $f$ is the median value of the template in that bin, normalized by the SED's $L_{\text{bol}}$, and $\bar{y}$ is the set of luminosity points within the bin.\\
For the lum-K13 template, we found that the scatter, $\sigma_{SED}/f$, increases as a function of the wavelength, ranging from 0.06 to approximately 0.2 for $\lambda$ between 1216 \AA\ and 1 \textmu m, and reaching up to 0.3 in the infrared at $\lambda \approx$ 5 \textmu m. In contrast, the WISSH-S23 template exhibits a more constant trend, with $\sigma_{SED}/f \approx$ 0.15 between $\lambda$ = 1216 \AA\ and 3 \textmu m, although with considerable fluctuations likely due to the small sample size. \\

For what concerns photometric points, we accounted for the contribution of emission lines to the observed magnitudes. Emission from the broad line region can contribute up to 30\% of the observed luminosity \citep[e.g.,][]{Miller:2023}. Therefore, when an emission line falls within a filter's transmission curve, it can lead to a significant overestimation of the QSO's primary emission. To estimate the contribution of emission lines, we employed an approach analogous to that used by \cite{Krawczyk:2013} and \cite{Saccheo:2023}. This involves determining the difference between the magnitudes obtained by convolving the composite spectrum from \cite{Vanden-Berk:2001} with the filters and those obtained by using only the underlying continuum. The resulting magnitude difference, $\Delta m$, which varies depending on redshift and the filter considered, is then subtracted from the observed magnitudes (i.e., resulting in larger corrected magnitudes).
To verify the consistency of this procedure, we repeated the analysis using the E-XQR-30 composite spectrum\footnote{https://github.com/XQR-30/Composites}. Since this spectrum was derived from QSOs in our sample, it is more representative of the spectral features of the sources analyzed. However, we found only minimal differences between the two approaches ($\Delta m_{XQR_{30}} - \Delta m_{\text{VB+01}} \lesssim 0.005$). Consequently, we chose to use the composite spectrum from \cite{Vanden-Berk:2001} as it extends beyond $\lambda$ = 3500 \AA\, allowing us to estimate the contributions from the H$\alpha$ and H$\beta$ lines as well. \\

SED fitting on the corrected photometric points was performed for each of the two templates by using the python package EMCEE \citep[][]{Foreman-Mackey:2019} to maximize the posterior arising from the likelihood 
\begin{equation}
\label{eqn:sed_fitting likekelihood}
   \mathcal{L}(K, E[B-V]) = -\sum_{i} \left(\frac{(y_{i}- K f_{i}10^{-0.4A_{\lambda}})^{2}}{\sigma_{i,tot}^{2}} + \log(2\pi \sigma_{i,tot}^{2}) \right)
\end{equation}
where $f_{i}$ denotes the flux obtained by convolving the SED template with the i-th filter transmission, $A_{\lambda}$ is the dust-reddening law by \cite{Prevot:1984} which accounts for a possible contribution by dust extinction \citep[see also][]{Bongiorno:2012} and $\sigma_{i, tot}^{2} = \sigma_{i,L}^{2} + \sigma_{i, SED}^{2}$ is the sum of the squared uncertainties on the luminosity points and the scatter of the template SED at that wavelength, which also depends on the normalizing constant $K$. Data points falling at $\lambda<$ 1216 \AA\ were not taken into account into the fit.
For both $K$ and $E[B-V]$, we assumed constant, non-negative priors and derived their best-fit values as the mean of the posterior distribution, with uncertainties taken from the 16th and 84th percentiles. 
We note that this SED-fitting procedure assumes that each QSO has an intrinsic continuum which is the same as the one of the employed template and any deviation from it is interpreted as the effect of dust extinction. Since it is possible for the QSO to be actually intrinsically redder than the average template \citep[see e.g.][]{Richards:2003, Krawczyk:2015} we advise caution particularly with very low measured E[B-V] values.\\

\subsection{Results}
\begin{figure}
    \centering
    \includegraphics[width = 0.48\textwidth]{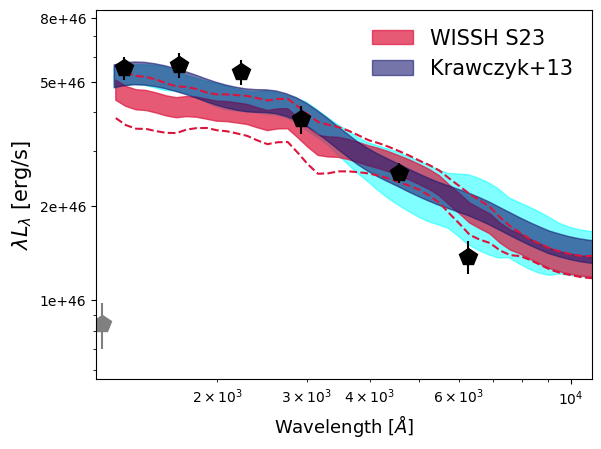}
    \caption{SED fitting results for J083+11.8 at $z = 6.346$ using lum-K13 (blue) and WISSH-S23 (red) templates. The thickness of the lines denotes the $\pm 1\sigma$ interval. Grey points denote photometry below the Ly$\alpha$. The cyan shaded area and the dashed red lines denote the $\pm$ 1$\sigma_{SED}$ region associated to the best fit for the lum-K13 and WISSH samples respectively, see Sec. \ref{sec:sed_fitting_uv}.}
    \label{fig:j083_detail}
\end{figure}

\begin{figure*}[h!]
    \centering
    \begin{tabular}{cc}
        \includegraphics[width = 0.40\textwidth]{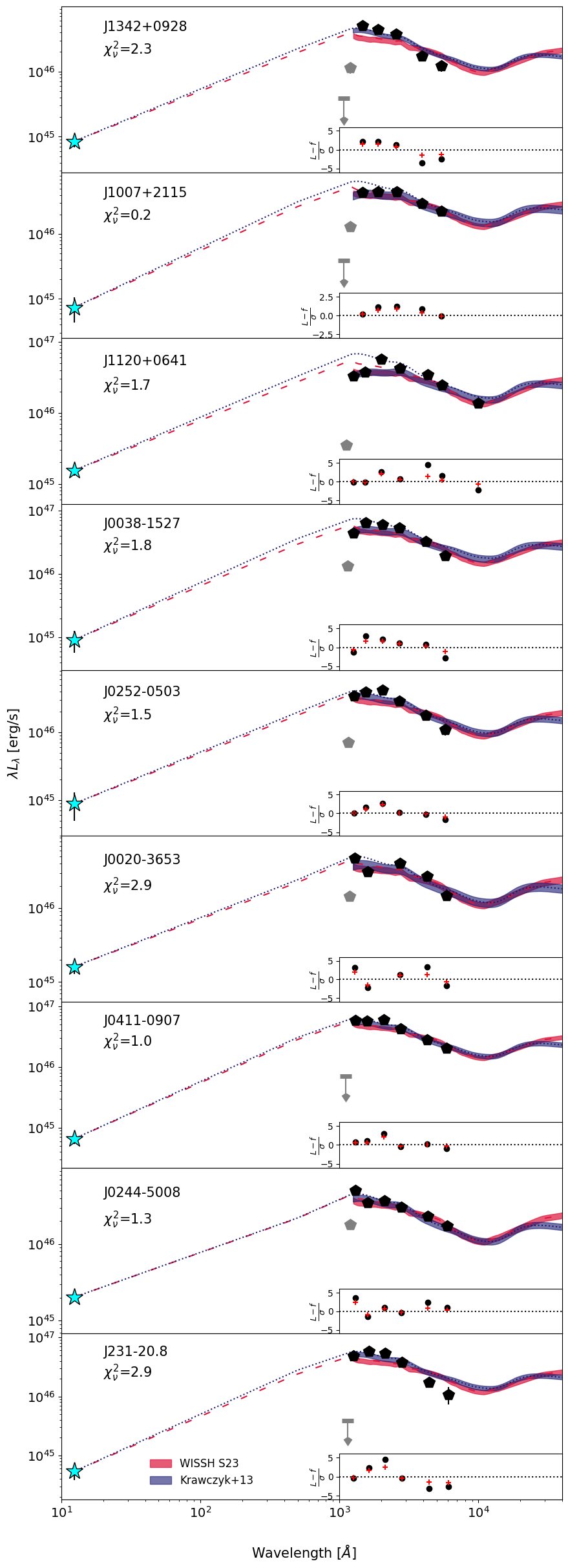} &
        \includegraphics[width = 0.40\textwidth]{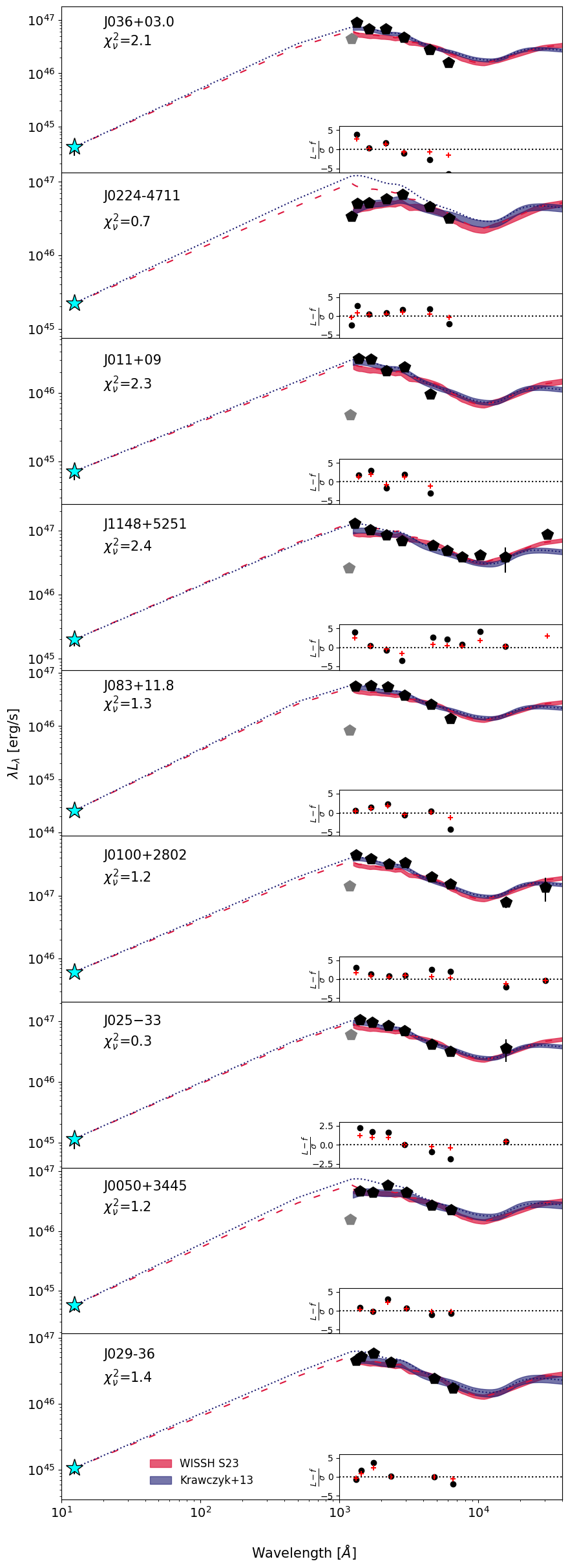} \\
    \end{tabular}
 \caption{Results of the SED fitting of the HYPERION QSOs using the lum-K13 (blue) and S23 (red) templates. The dotted blue and dashed red lines give the unextincted SEDs which have been prolonged below the Ly$\alpha$ with a double power-law as explained in Sect. \ref{sec:emphirical_fitting}; $L_\text{bol}$ is the integral under the dotted/dashed lines. Data at UV wavelengths shorter than Ly$\alpha$ were not considered in the fitting and are colored gray.
 The cyan stars indicate measured 1 keV luminosities, see Tab. \ref{table:luminosities}. The sub-panels show the residuals with respect to the K13 best-fit SED. The black circles refer to the residuals computed using only the uncertainties on the photometric points while red crosses represent residuals for which also the scatter on the SED template was taken into account.}
\label{fig:fitting}
\end{figure*}

We find that, for 73\% of our sources, the lum-K13 template achieves better results, in $\chi^{2}_{\nu}$ terms, in describing the broad band SED of our QSOs, with a median $\chi^{2}_{\nu}$ = 1.3 compared to $\chi^{2}_{\nu}$ = 1.8 obtained with WISSH-S23. For this reason, throughout the paper,  all values reported (i.e. $L_\text{bol}$, E[B-V]), are derived from the fitting obtained with the lum-K13 template.
Examining in detail the SEDs (e.g. J083+11.8, shown in Fig. \ref{fig:j083_detail}), we notice that the WISSH-S23 template yields worst results due to a flatter UV continuum for $\lambda < $ 2000 \AA\ a feature not found in most of our EoR QSOs. \\
The HYPERION QSO SEDs with the final best fits for both lum-K13 and WISSH-S23 templates are reported in Fig. \ref{fig:fitting} while Fig. \ref{fig:xqr_30_sed} shows the modeling with lum-K13 for the E-XQR-30 sample. 
Upon visually inspecting the fitted SEDs, we observe that the data points with $\lambda < 1$ \textmu m are generally well-fitted, while those at longer wavelengths exhibit larger discrepancies from the model (see also Sect. \ref{sec:NIR}). This outcome is expected, as previously discussed, QSOs tend to show greater scatter around their mean value in the NIR region.\\
Two notable exceptions with particularly poor fits in the UV-optical region, although their $\chi^{2}_{\nu}$ are not the worst, are PSOJ023-02 and SDSSJ0836+0054. PSOJ023-02 has a very flat SED, which is not well modeled by reddening and may be intrinsic in nature, as it does not display the typical curling of the SED caused by dust extinction \citep[i.e the reddening it stronger at shorter wavelengths, see][]{Hopkins:2004}. In contrast, the fit for SDSSJ0836+0054 is quite accurate for most points, except for the Spitzer $S_{4.5}$ data point, which is significantly more luminous than the modeled SED. While we cannot definitively explain this discrepancy, a plausible reason could be that the $S_{4.5}$ flux is overestimated. This is supported by the fact that, although probing almost the same wavelength range, the value reported for the W2 filter is 0.28 mag fainter.\\

\label{sec:emphirical_fitting}

\subsection{X-ray and EUV SED modeling}
To model the SEDs in the unobserved Extreme UV (EUV, 12.4  $\leq \lambda/\AA <$ 1216 ) we followed the method discussed in \cite{Lusso:2012} and \cite{Shen:2020}, i.e. a power-law with a fixed slope $\lambda L_{\lambda} \propto \lambda^{0.8}$ in the range between 500 $\leq \lambda /$\AA $\leq$ 1216, derived from HST observations of local AGN (\citealt{Zheng:1997}, \citealt{Telfer:2002}, \citealt{Lusso:2015}), plus a power-law with a free-to-vary spectral slope $\alpha$ to link $L_{500\text{\AA}}$ with the 1 keV luminosity. Thus, the full SED template takes the following form:

\begin{equation}
\begin{cases}
f^{SED}(\lambda)\; \text{at}\; \lambda > 1216\,{\text{\AA}},\;  f^{SED}(\lambda) =\text{lum-K13 or WISSH-S23} \\
\lambda L_{\lambda} \propto \lambda^{0.8}\; \text{at}\; \;500\,{\text{\AA}}<\lambda \leq 1216\,{\text{\AA}}\\
\lambda L_{\lambda} \propto \lambda^{-(\alpha+1)}\;\text{at}\;  12.40\,{\text{\AA}}<\lambda \leq 500\,{\text{\AA}}\\
\end{cases}
\label{eqn:double_PL}
\end{equation}
The derived $\alpha$ for HYPERION QSOs are reported in Tab. \ref{table:spectral_slopes}.
The same is valid for the E-XQR-30 QSOs but, since there are not 1 keV luminosities, we employed the value obtained as the geometric mean of the HYPERION sample, i.e $Log(\lambda L_{1kev}) = 45.01 \pm 0.08$ erg/s. The advantage of using a double power-law to describe EUV emission lies in the fact that it allows us to make use of both the information available on the QSOs emission just below the Ly$\alpha$ line (from low-z sources) and the high-quality X-ray data obtained for these sources at \textit{z} $\geq$ 6. This would not have been possible using a fixed template or extrapolating with a single power-law.  
\begin{figure}[h]
    \centering
    \includegraphics[width=0.48\textwidth]{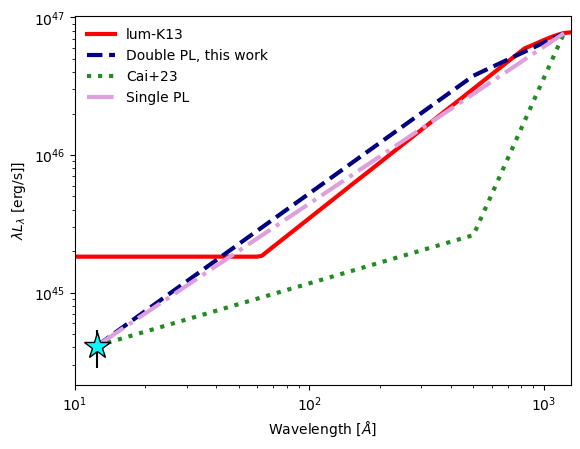}
    \caption{Different parametrizations considered for the EUV modeling of the SEDs applied to J036.5+03.0, see Sect. \ref{sec:emphirical_fitting}.}
    \label{fig:euv_parametrization}
\end{figure}
Since the EUV contributes to nearly 50\% of the QSO emission, different modelings could lead to significantly different derived $L_\text{bol}$, which is the main value we want to derive. 
Therefore, we tested  the impact of different modelings on the integrated luminosity under the EUV region.
Our double power-law parametrization yields, on average, EUV luminosities $\sim$ 20\% larger than those derived using the EUV modeling of lum-K13 with differences up to 40\%. Conversely, employing a single power law extrapolation yields an average decrease of 8\%, with a maximum deviation of 38\%.
Finally, by employing the far redder EUV spectral slope recently proposed by \cite{Cai:2023}
we obtain EUV luminosities which are 70\% lower compared to ours.
Fig. \ref{fig:euv_parametrization} shows these three different modelings for the case of J036+03.0 which, in terms of optical-to-X-ray ratio, can be considered as representative of the entire HYPERION sample \citep[see][]{Zappacosta:2023}.

We have also tried to quantify the possible uncertainties associated to the fact that for E-XQR-30 sources we assumed an average $L_{1keV}$ calculated from the HYPERION sample.  
Indeed, although the proper X-ray emission, i.e. $\sim 20\, \text{\AA} - 1$ keV only accounts for less than 1\% of $L_\text{bol}$ in these luminous QSOs \citep[e.g.][]{Duras:2020},  the 1 keV luminosity constrains the slope of the second
power-law in our modeling of the unseen EUV SED, and thus may potentially affect $L_\text{bol}$.
To quantify the impact of this assumption, we simulated a QSO with $log(L_{2500\text{\AA}})$ = 47 erg/s and we calculated how much $L_\text{bol}$ varied in the case that it had an X-ray luminosity twice (half) of that predicted by the relation by \cite{Lusso:2010}, finding an increase (decrease) in  $L_\text{bol}$ of  $\sim$ 4\%.

\section{Bolometric luminosities and E[B-V]}
The bolometric luminosities for the HYPERION and E-XQR-30 QSOs have been computed by integrating their dereddened SEDs obtained via lum-K13 template between 1 keV and 1 \textmu m, where the choice of integration limits ensures to not include reprocessed radiation and thus, to avoid counting the same contribution twice \citep[see][]{Marconi:2004}. Assuming that $L_\text{bol}$ were computed without dereddening the SED, we would obtain lower values for $L_\text{bol}$, with a difference $\Delta \log(L_\text{bol}) \approx 4 \times E[B-V]$, which is nearly linear within the range of $E[B-V]$ explored by our sources.\\
 $L_\text{bol}$ values are provided in Tab. \ref{table:luminosities} (HYPERION sources) and Tab. \ref{table:xqr30_luminosities} (E-XQR-30 sources) as well as monochromatic luminosities at various wavelengths obtained by interpolating two adjacent photometric points and corrected for the SED-fitting derived $E[B-V]$ of the source. We also calculated $\lambda_\text{Edd}$ assuming the $M_\text{SMBH}$ derived from the MgII line single epoch virial mass estimator \citep[see Tab. \ref{table:coordinates} and][for further references and for the associated uncertainties]{Zappacosta:2023, Mazzucchelli:2023}.
\begin{figure*}[]
    \sidecaption
    \includegraphics[width=12cm]{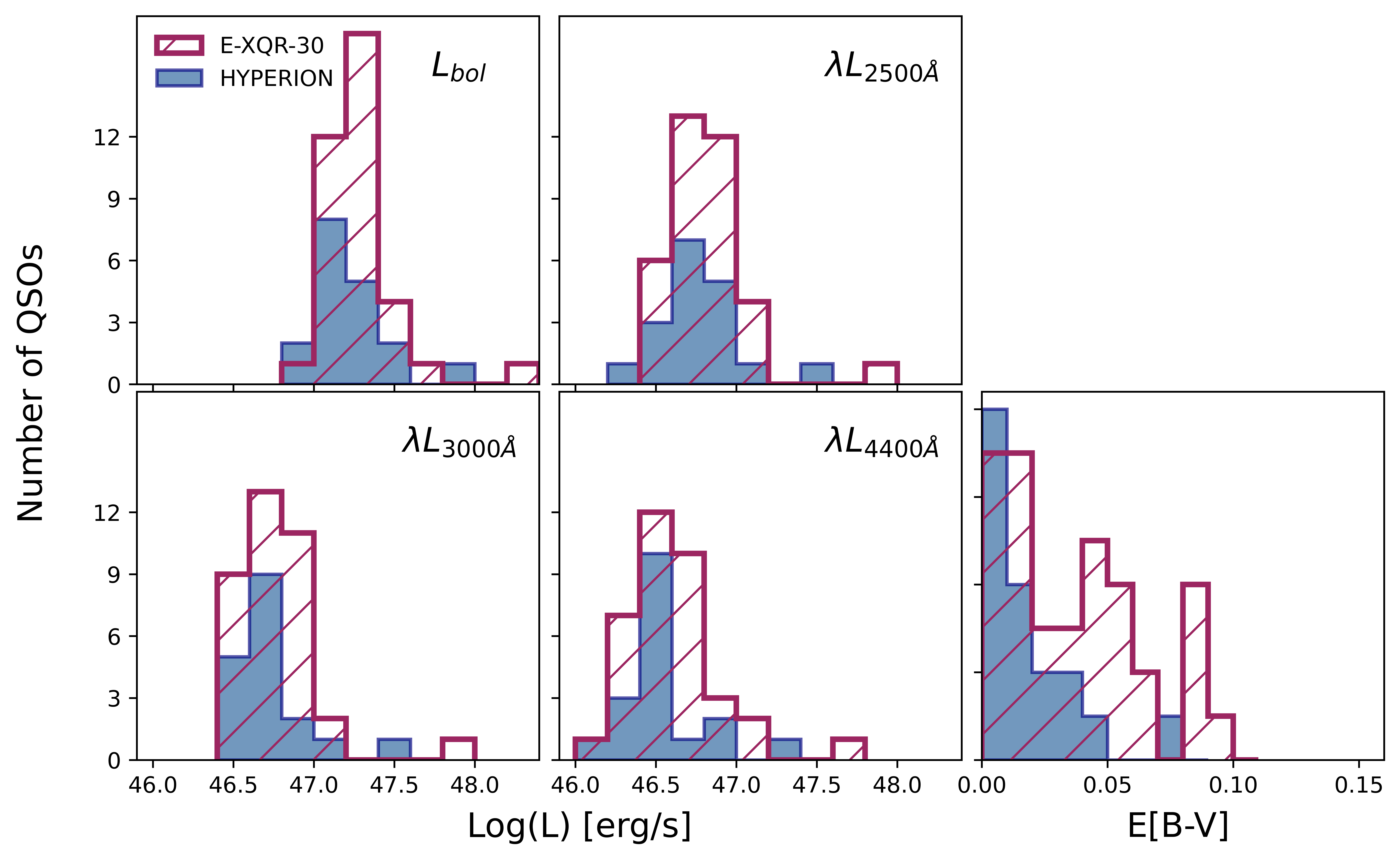}
    \caption{Bolometric and monochromatic luminosities and E[B-V] distributions for the HYPERION (blue) and E-XQR-30 (purple) samples.}
    \label{fig:luminosities_histograms}
\end{figure*}

In Fig. \ref{fig:luminosities_histograms} we show the luminosity distributions of both HYPERION and E-XQR-30 samples, which range  between $46.5 \lesssim \log(L_\text{bol}/[erg/s]) \lesssim 48.1$;
performing a Kolmogorov-Smirnov test on the $L_\text{bol}$ distributions of the two samples gives a p-value of $\sim$ 0.05 (i.e. it does not reject the hypothesis that both samples have been drawn from the same distribution), thus strengthening our choice of using the E-XQR-30 QSOs as a complementary sample to HYPERION.\\
\begin{figure}
    \centering
    \includegraphics[width = 0.49\textwidth]{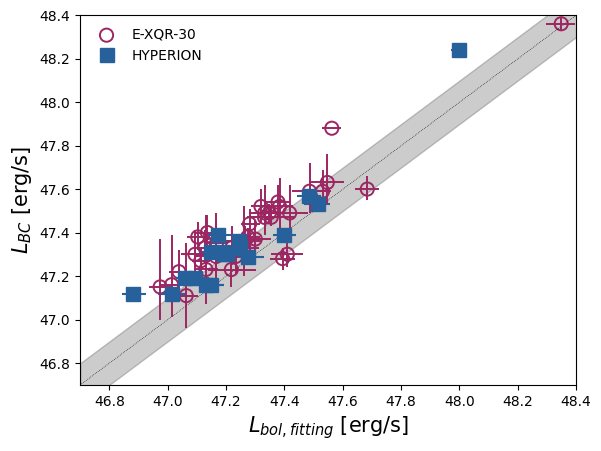}
    \caption{$L_\text{bol}$ already published \citep[][]{Zappacosta:2023, Mazzucchelli:2023} vs $L_\text{bol}$ computed in this work. $L_\text{bol}$ in the literature have been computed using the 3000 \AA\ BC ($L_\text{bol} = 5.15 L_{3000\text{\AA}}$). The black line gives the 1:1 ratio while the grey shaded area delimits the 0.1 dex difference region.}
    \label{fig:Lbol_BC}
\end{figure}

In Fig. \ref{fig:Lbol_BC} we compare $L_\text{bol}$ with those computed via BC, $L_\text{BC}$, \citep[][and references therein]{Mazzucchelli:2023, Zappacosta:2023}. We find that, when integrating the full SED, we get systematically lower values with, on average, $\langle\Delta(L_\text{bol}) \rangle = 0.13 \pm 0.08$ dex (i.e. $L_\text{BC}$ are overestimated by a factor of 34\%), and discrepancies up to $\sim$ 0.25 dex in few QSOs. However, we notice that $L_{BC}$ were derived using the 3000 \AA\ bolometric correction $BC_{3000\text{\AA}} = 5.15$ \citep[][]{Richards:2006, Shen:2011} which is obtained by integrating the entire SED between 10 keV and 100 \textmu m. Therefore, including also X-ray and IR reprocessed emission leads to the overestimation of $L_\text{bol}$ compared to our values.
This overestimation also impacts the determination of $\lambda_\text{Edd}$, which we find, on average, to be 0.12 $\pm$ 0.06 dex lower than the previous measurement.\\
Regarding the derived E[B-V], the distribution is dominated by sources not affected by dust reddening, i.e. with E[B-V] = 0, see right panel in Fig. \ref{fig:luminosities_histograms}. However we notice a slight disagreement between the HYPERION and E-XQR-30 samples; while only J0224-4711 among the HYPERION QSOs has E[B-V]$\geq$ 0.05, there are 11 E-XQR-30 sources with E[B-V] values exceeding this value (excluding J0224-4711). Given the relation between dust extinction and the presence of BAL features \citep[][]{Gallagher:2007, Bischetti:2022} this difference can be accounted for the fact that, while broad absorption lines are found in 47\% of E-XQR-30 sources, 
only J0038-1527 \citep[][]{Wang:2018} and J231-20 \citep[][]{Bischetti:2022} have published BAL features among HYPERION QSOs, (although, to date, no BAL feature analysis has been performed as systematically and consistently as done for E-XQR-30 QSOs.) A potential caveat regarding the correlation between BAL troughs and increased dust extinction is that QSOs may appear reddened not due to actual reddening, but because their intrinsic fluxes are underestimated in regions of the spectrum affected by absorption features. However, while this could indeed be true for low measured E[B-V] values ($\sim 0.01-0.03$), the absorption features typically observed in the E-XQR-30 sample are not deep enough to influence flux measurements to the extent that they could mimic the effects of significant E[B-V]. Moreover, \cite{Bischetti:2022} already noted that BAL QSOs in the E-XQR-30 sample exhibit redder W1-W2 colors, which are bands that do not probe wavelengths affected by absorption troughs.

\subsection{Bolometric corrections}
\begin{figure*}
    \sidecaption
    \includegraphics[width = 12cm]{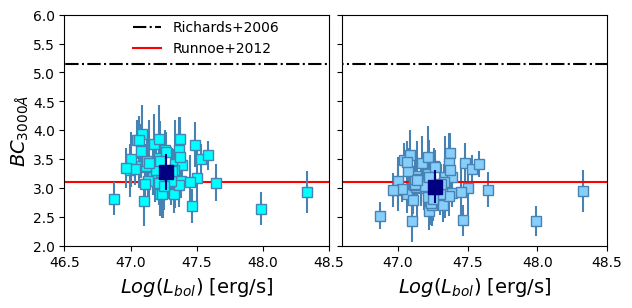}
    \caption{$BC_{3000\text{\AA}}$ \textit{vs} $L_\text{bol}$  for HYPERION and E-XQR-30 QSOs with their weighted mean value reported as a dark blue square. The left panel illustrates $L_\text{bol}$ computation using equation (\ref{eqn:double_PL}), while the right panel showcases results obtained through the lum-K13 modeling of the EUV.
     The constant 3000 \AA\ BCs proposed by \cite{Richards:2006} are overplotted. The original value of 5.15 is depicted as a black dot-dashed line while the version recomputed by \cite{Runnoe:2012} to exclude IR emission is represented as a red solid line.}
    \label{fig:BC_3000}
\end{figure*}

\begin{figure*}[h]
    \sidecaption
    \includegraphics[width = 12cm]{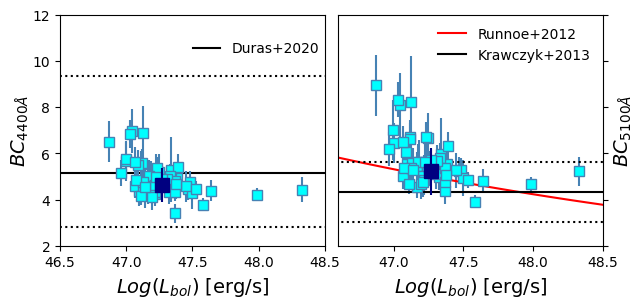}
    \caption{Bolometric corrections at 4400 (\textit{left}) and 5100 \AA\ (\textit{right}) with their weighted mean values reported as blue squares. Overplotted are the relationships by \cite{Duras:2020}, \cite{Krawczyk:2013} (with the associated uncertainties as dotted lines) and \cite{Runnoe:2012}.}
   \label{fig:bolometric_corrections}
\end{figure*}

The measurement of $L_\text{bol}$ for high-z QSOs usually relies on the application of a bolometric correction, especially the  3000 \AA\ one. Since the used values have been derived from lower-\textit{z} sources  we investigated whether these values are appropriate for objects at the EoR  or whether it is necessary to use different values.
In Fig. \ref{fig:BC_3000}, left, we show our derived $BC_{3000\text{\AA}}$ for the HYPERION and E-XQR-30 QSOs compared to the constant bolometric correction by \cite{Richards:2006}, i.e. $BC_{3000\text{\AA}}$ = 5.15 which has been adopted in the literature as a standard $BC_{3000\text{\AA}}$ for z$>$6 QSOs \citep[e.g] []{Wu:2015, Banados:2018, Reed:2019,Shen:2019, Yang:2021, Farina:2022, Mazzucchelli:2023, Fan:2023, Zappacosta:2023}. We also show a re-evaluation of this value by \cite{Runnoe:2012} to exclude reprocessed IR emission which allows for a fair comparison with our values, i.e., $BC_{3000\text{\AA}} = 3.11$.
The $BC_{3000\text{\AA}}$ distribution of  HYPERION and E-XQR-30 sources ranges between 2.64 and 3.94, with a mean value of 3.30 $\pm$ 0.3. The majority of derived BCs are higher than the recalculated $BC_{3000\text{\AA}}$ by \cite{Runnoe:2012} although still in full agreement within the uncertainties. This small discrepancy, rather than arising from a different SED shape in our objects, can be attributed to our more careful modeling of the EUV (i.e. with the double power law) which, as mentioned above, results in higher $L_\text{bol}$ (and consequently higher BC) than those obtained using an average SED template. The right panel of Fig. \ref{fig:BC_3000} illustrates that when computing $L_\text{bol}$ with the lum-K13 EUV recipe, the BCs are fully consistent with the predictions by \cite{Runnoe:2012}.\\
Given our more refined treatment of the UV-X-ray SED
which gives more accurate $L_\text{bol}$ measurements, we recommend using the revised  $BC_{3000\text{\AA}}$ value instead of 5.15 to recover a more accurate $L_\text{bol}$ for z$>$6 QSOs. This will further allow to remove a source of inaccuracy in the determination of $\lambda_\text{Edd}$ which is already affected by systematics from the mass determination.\\
Additionally, we computed optical bolometric corrections at 4400 \AA\ and 5100 \AA, comparing them to relationships established for lower-\textit{z} sources, as depicted in Fig. \ref{fig:bolometric_corrections}. For $\lambda = 4400$ \AA, the bolometric corrections demonstrate a good agreement with the average value reported in \cite{Duras:2020} for a large collection of AGN at 0 $\lesssim z \lesssim$ 3, resulting in a mean of 4.9 $\pm$ 0.7. In the case of 5100 \AA\ BC, our derived $BC_{5100\AA}$ values are consistent with the extrapolated predictions at higher luminosities from the relationship provided by \cite{Runnoe:2012}. However, they almost all exceed the value reported in \cite{Krawczyk:2013}, although their mean value $BC_{5100\text{\AA}} = 5.7 \pm 0.9$ remains in agreement within 2$\sigma$.

\section{Predicting X-ray luminosity with a physically motivated AD model}
\label{sec:qsosed}

\begin{figure*}[h]
    \centering
    \begin{tabular}{cc}
        \includegraphics[width = 0.40\textwidth]{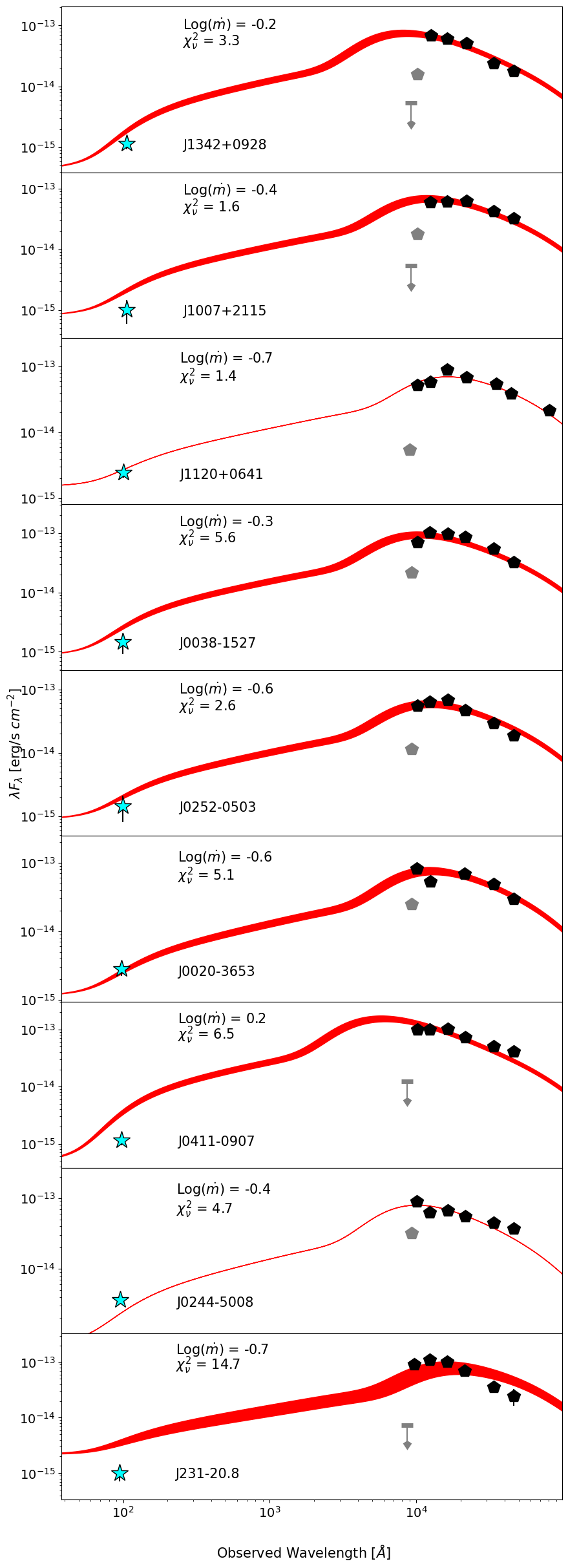} &
        \includegraphics[width = 0.40\textwidth]{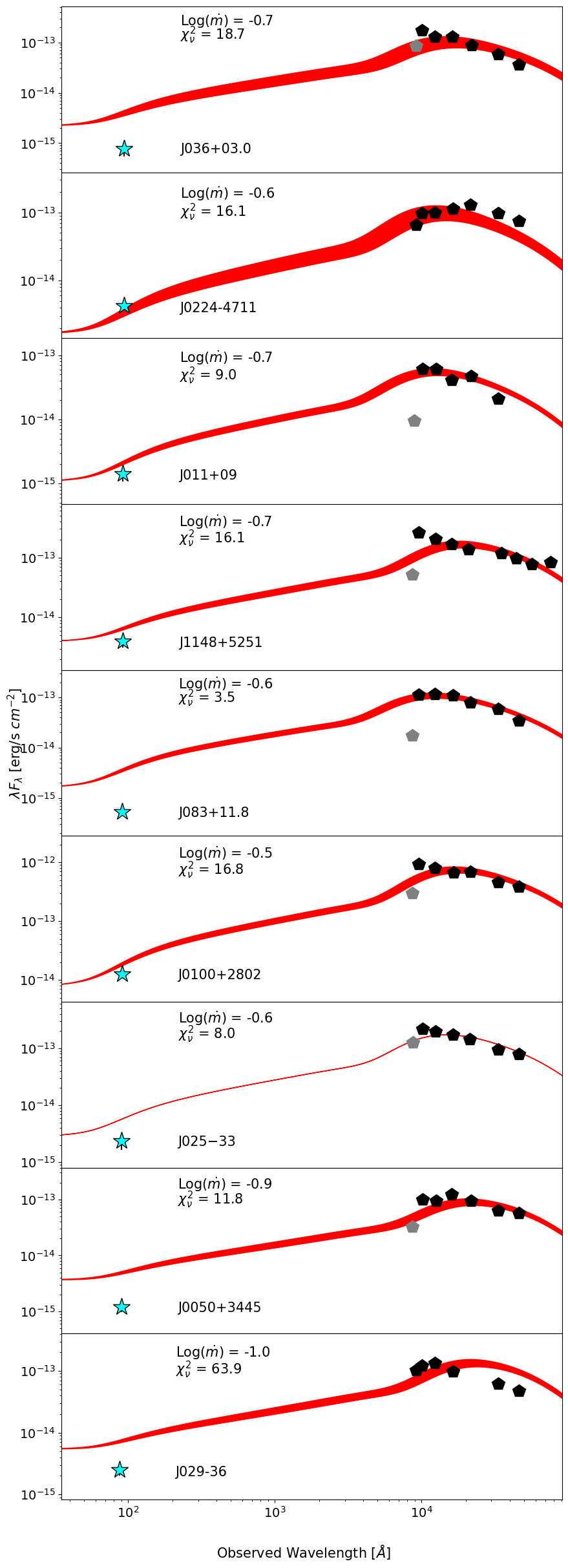} \\
    \end{tabular}
 \caption{Same as in Fig \ref{fig:fitting} with QSOSED generated templates. Each panel presents the mass accretion and the $\chi^{2}_{\nu}$ derived by the fit.}
\label{fig:fitting_xspec}
\end{figure*}

We made use of QSOSED \citep[][]{Kubota:2018}, a physically-motivated model for disc/corona emission, to describe the nuclear emission of QSOs. Specifically, we aimed to determine whether these models can accurately predict the coronal emission (i.e., the 1 keV luminosities) by fitting only the accretion disk data, which includes the rest-frame UV and optical photometric points. Accordingly, our analysis was limited to the 18 HYPERION QSOs.
QSOSED  assumes that the emitted radiation originates from three distinct regions at increasing radial distances: (i) a hot ($k_{B}$T $\gtrsim$ 10 keV) comptonizing plasma \citep[i.e. the corona, e.g.][]{Haardt:1991} at $R_\text{ISCO} < r \leq R_\text{hot}$, which is responsible for the power-law like X-ray continuum emission, (ii) a warm ($k_{B}$T $\lesssim$ 5-10 keV) comptonizing plasma in an intermediate region $R_\text{hot} < r \leq R_\text{warm}$ contributing to the soft excess \citep[][]{Petrucci:2018}, and, (iii) a standard, geometrically thin, optically thick, accretion disk at $r > R_\text{warm}$. Furthermore QSOSED assumes that the hot corona radiates 2\% of the Eddington luminosity of a source, regardless from its $L_\text{bol}$.  \\
SED templates calculated by QSOSED were generated through the X-ray spectral fitting program XSPEC \citep[][]{Arnaud:1996} by taking into account four main free parameters, namely $M_\text{SMBH}$, the BH spin, the specific mass accretion rate, i.e. $\dot{m}$, and the cosine of the inclination angle $i$, i.e. $\cos(i)$. Given the high probability of resulting in degenerate templates, due to the relatively low number of degrees of freedom (i.e. photometric points), we fixed the BH spin to 0, $M_\text{SMBH}$ to the values listed in \cite{Zappacosta:2023} and the viewing angle to 30$^{\circ}$ which is an average value among those expected for type-1 QSOs \citep[e.g.][]{Mountrichas:2021}. Accordingly the only free-to-vary parameter was $\log(\dot{m})$ which ranges from -1 to 0.4 in steps of 0.1. 
Furthermore, we did not account for dust reddening since we derived from the empirical modeling that is negligible for the vast majority of the sources (see  Tab. \ref{table:luminosities}).
In QSOSED, the luminosity from the QSO and thus the observed flux at each wavelength are predetermined without any normalization. Given the many assumptions on our input parameters, and the systematic uncertainties on $M_\text{SMBH}$, \citep[ $\sim$ 0.55 dex, e.g.][]{Mazzucchelli:2023}, we expect a limited accuracy in the SED description which, we emphasize, is not the main result we aim to achieve with this analysis.
In this model, we cannot continuously explore the parameter space, as the allowed values of $\dot{m}$ are discretized. Thus, the best fit was determined via $\chi^{2}_{\nu}$ minimization, while uncertainties were estimated by identifying the range of $\dot{m}$ that resulted in $\Delta \chi^{2}_{\nu} \leq 1$ from the best fit, corresponding to a 1$\sigma$ confidence interval \citep[e.g.,][]{Avni:1976}.\\
The results from the fits, shown in Fig. \ref{fig:fitting_xspec} together with the derived $\dot{m}$, indeed reveal a quite large range of $\chi^{2}_{\nu}$ values, varying from 1.4 for J1120+0641 to $\sim$ 20, with J029-26 being a notable outlier with $\chi^{2}_{\nu}=77$ and a $\chi^{2}_{\nu}$ median value of 7.2.
Interestingly, among the 13 QSOs for which the QSOSED template provided an acceptable fit (arbitrarily chosen to be those with $\chi^{2}_{\nu}\leq 15$) the 1 keV fluxes predicted by QSOSED are consistent with the measured ones within $3\sigma$ for 7 sources (54\%) and within 5$\sigma$ for 9 sources (70\%). By examining the difference in the ratio of predicted to measured fluxes, QSOSED models deviate by a median factor of 1.6, with a maximum discrepancy factor of 6.3 in the case of J083+11.8.
This outcome highlights the effectiveness of QSOSED in predicting X-ray emission based solely on the observed luminosity from the accretion disk even in case of the most distant QSOs. 
Noteworthy, for the three sources showing  a significant difference between observed and  QSOSED predicted X-ray fluxes (namely, J0411-0907, J083+11.8, and J0050+2445), the observed X-ray flux is always lower than the predicted one by factors of 2.7, 6.3 and 3.8 respectively.\\
The deviation of the QSOSED prediction at 1 keV from the actual data may reflect either (i) the deviations from intrinsic physical properties (e.g. mass measurement, spin assumption, source inclination and lack of extinction) for some sources or (ii) the underlining assumptions on which this model is based, i.e. fraction of the X-ray coronal emission in terms of $L_\text{Edd}$, $R_\text{hot}$ and $R_\text{warm}$, temperature of the hot and warm Comptonizing regions. An investigation of the cause of the disagreement is beyond the aim of this paper and it is deferred to a future work. \\
Finally, in light of the good predictive results demonstrated by QSOSED, we performed the fitting under the same assumptions (i.e. $i=30^{\circ}$, spin =0) on E-XQR-30 QSOs to estimate their 1 keV fluxes,  which are reported, converted to luminosities, in Tab. \ref{table:xqr30_luminosities}.

\section{UV and optical slopes}
\label{sec:slopes}
\begin{figure}[h!]
    \centering
    \includegraphics[width = 0.49\textwidth]{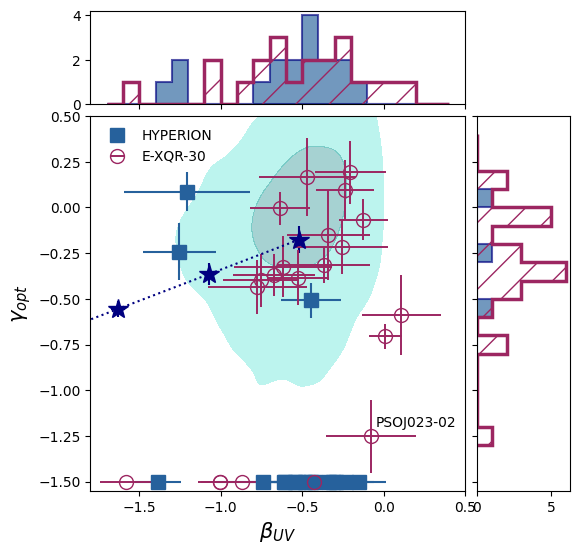}
    \caption{$\beta_\text{UV}$ (1300 $<\lambda/\AA<$ 3000) and $\gamma_\text{opt}$ ((3000 $<\lambda/\AA<$ 10000) distributions for the sample. The points along the constant line $\gamma_\text{opt}$ = -1.5 have only the UV slope measured. Blue stars represent the values obtained from lum-K13 template assuming E[B-V] values of 0, 0.05 and 0.1 assuming observations in the UKIDSS YJHK bands plus W1 and W2 for QSOs at 5.5 $\geq z \geq$ 7.5. The shaded light blue areas delimit the 1 and 2 $\sigma$ confidence intervals for the luminous QSO subsample in \cite{Krawczyk:2013}. }
    \label{fig:slopes}
\end{figure}

The optical-UV SED is usually modeled as a broken power law, with the break between the two slopes falling at $\lambda = 3000-5000$ \AA\ \citep[e.g.][]{Vanden-Berk:2001, Temple:2021b}. 
To derive the slopes of the HYPERION and E-XQR-30 QSOs we opted to follow \cite{Lusso:2016}, i.e. we set the break at $\lambda = 3000\;$ \AA\ although we extended the UV interval down to 1300 \AA\ instead of 1450 \AA\ in order to include for more QSOs the Y band photometric point, without being affected by the peak of the Ly$\alpha$ emission. Therefore we refer to $\beta_\text{UV}$ as the UV slope, characterizing the accretion disk emission between $\lambda$ = 1300 \AA\ and 3000 \AA\ and $\gamma_\text{opt}$ as the optical slope, describing the emission at longer wavelengths, up to $\lambda$ = 1.0 \textmu m, where the SED shows an inflection due to the arising contribution from hot dust. Spectral slopes are reported in luminosity density units, i.e. in the form $L_{\nu} \propto \nu^{\beta}$.\\
Slopes were computed independently via least-square minimization by fitting with a straight line the photometric points in the log-log space. To compute the slope we required the QSOs to have at least three photometric points in the interested wavelength interval; this requirement limited the number of QSOs with measured $\beta_\text{UV}$ and $\gamma_\text{opt}$ to 39 and 30 respectively. In particular, all but three sources with computed $\gamma_\text{opt}$ belong to the E-XQR-30 sample as they have lower redshifts (indeed at $z \gtrsim$ 6.3 the K band moves at shorter wavelengths than 3000 \AA). Fig. \ref{fig:slopes} shows the computed $\beta_\text{UV}$ and $\gamma_\text{opt}$ distribution (reported in Tab. \ref{table:spectral_slopes} and Tab. \ref{table:xqr30_spectral_slopes}). For comparison it is also shown the contour enclosing the 39 and 86.4\% (respectively 1 and 2$\sigma$ under the assumption of a bivariate normal distribution) of the values derived from the \cite{Krawczyk:2013} luminous subsample following an analogous methodology. Additionally, since the derived  slope could vary as a function of \textit{z} simply because the filters are sampling different regions of the SED we also overplot the mean $\pm 1\sigma$ values obtained by redshifting the lum-K13 template at 5.5 $\geq z \geq $ 7.5 and generating mock observations with UKIDSS YJHK plus W1 and W2 filters for different levels of E[B-V].\\
The HYPERION and E-XQR-30 sources are in complete agreement with the ranges of values observed in luminous QSOs at lower \textit{z}, and, considering the uncertainties, which are quite large  due to the fitting being performed on a limited number of points the bulk of the sample agrees with the point obtained from the lum-K13 template without dust extinction. 
QSOs having lower $\beta_\text{UV}$ (i.e. a flatter SED) can be explained through dust reddening and indeed they are generally the ones having higher E[B-V] values. In particular the three HYPERION sources with lower $\beta_\text{UV}$ are those with measured E[B-V] $\geq$ 0.03 (see Tab. \ref{table:luminosities}). 
Conversely, two E-XQR-30 QSOs exhibit bluer $\beta_\text{UV}$ than the rest of the population although still within $\sim 2\sigma$ from the distribution observed in \cite{Krawczyk:2013} while, PSOJ023-02, located in the bottom-right of Fig. \ref{fig:slopes} shows a $\gamma_\text{opt}$ value in significant disagreement with the rest of the distribution. 
As already mentioned in Sect. \ref{sec:sed_fitting} this object has a rather flat spectrum and hence the low $\gamma_\text{opt}$ is expected. 
Finally, we tested for potential correlations between $\beta_\text{UV}$ and both the photon index $\Gamma$ and the UV-to-X-ray ratio $\alpha_{OX}$, defined as $-0.384\log \left (L_{\nu, 2keV}/L_{\nu, 2500\AA}\right)$ and reported in \cite{Zappacosta:2023} and \cite{Tortosa:2024b}. However, we found no evidence of correlation among these parameters.
Indeed we simulated multiple realizations of the datasets by generating synthetic  values normally distributed around the best-fit value and with a $\sigma$ equal to their uncertainties and, performing Spearmann tests for each of them, obtained mean correlation coefficients and p-values of 0.22 and 0.4 for the $\Gamma$ vs $\beta_\text{UV}$ relation and -0.05 and 0.6 for the $\alpha_{OX}$ vs $\beta_\text{UV}$ one.

\begin{table}
\centering
\begin{tabular}{lccc}
\toprule
Name & $\alpha_{EUV}$	 & $\beta_{UV}$ & $\gamma_{opt}$ \\
\midrule
J1342+0928	&	-1.88	&	-0.15 $\pm $ 0.17	&	 \\
J1007+2115	&	-2.01	&	-0.74 $\pm $ 0.29	&	 \\
J1120+0641	&	-1.83	&	-1.21 $\pm $ 0.39	&	0.09 $\pm $ 0.11 \\
J0038-1527	&	-2.0	&	-0.61 $\pm $ 0.22	&	 \\
J0252-0503	&	-1.84	&	-0.42 $\pm $ 0.25	&	 \\ 
J0020-3653	&	-1.73	&	 	&	                 \\
J0411-0907	&	-2.04	&	-0.55 $\pm $ 0.15	&	 \\
J0244-5008	&	-1.65	&	-0.44 $\pm $ 0.12	&	\\ 
J231-20.8	&	-2.07	&	-0.31 $\pm $ 0.25	&	 \\
J036+03.0 	&	-2.21	&	-0.23 $\pm $ 0.13	&	 \\
J0224-4711	&	-1.89	&	-1.38 $\pm $ 0.14	&	 \\
J011+09 	&	-1.81	&	-0.52 $\pm $ 0.11	&	 \\
J1148+5251	&	-1.93	&	-0.45 $\pm $ 0.19	&	-0.51 $\pm $ 0.10 \\
J083+11.8	&	-2.28	&	-0.61 $\pm $ 0.14	&	 \\
J0100+2802	&	-1.94	&	-0.34 $\pm $ 0.11	&	 \\
J025-33	&	-2.02	&	-0.48 $\pm $ 0.09	&	      \\
J0050+3445	&	-2.12	&	-1.26 $\pm $ 0.22	&	-0.24 $\pm $ 0.16 \\
J029-36 	&	-1.9	&	-0.29 $\pm $ 0.16	&	 \\

\bottomrule
\end{tabular}
\caption{EUV (12.4 $\leq \lambda/\AA \leq$ 500, not fitted), UV  (1300 $\leq \lambda/\text{\AA} \leq$ 3000) and optical  (3000 $\leq \lambda/\text{\AA} \leq$ 11000) derived spectral slopes for the HYPERION sample. }
\label{table:spectral_slopes}
\end{table}

\section{NIR modeling of the hot dust component}
\label{sec:NIR}
Although the available MIR points (Spitzer MIPS 24 \textmu m, W3, and W4) were not utilized in the fitting process, it is intriguing to compare their positions to what is expected from the normalized SED templates. Previous studies \citep[e.g.,][]{Jiang:2006, Leipski:2014, Bosman:2023} have reported a NIR emission of QSOs at the EoR resembling that of luminous objects at lower \textit{z}, characterized by a black-body-like continuum originating from a hot dust component close to its sublimation temperature. However, there are several indications suggesting a higher fraction of dust-poor objects at the EoR, with a lower-than-average NIR-to-optical emission ratio \citep[][]{Jiang:2006, Leipski:2014} or even dust-free sources \citep[][]{Jiang:2010}. In these cases, the NIR can be well-modeled by assuming only emission from the accretion disk, modeled as a power-law, without the need to add emission from hot dust. The cause of the observed dust poorness in these QSOs remains uncertain. It could be attributed to an actual deficiency of dust, possibly linked to these sources being in the early stages of the Universe \citep{Jiang:2010}. Alternatively, it may be associated with a distinct torus structure, characterized by a lower covering factor, resulting in less dust being directly exposed to the primary radiation \citep[see][]{Lyu:2017}.\\
In total,  we have 11 QSOs where the 24 \textmu m or the W4 photometry, necessary to constrain the hot dust emission, is available. We find that in 4 of them, lum-K13 SED agrees with the true NIR luminosity, while 5 objects have enhanced NIR emission and 2 objects have substantially lower emission.\\
To better quantify the strength of hot dust emission, following the approach by \cite{Temple:2021}, we computed for our sources $X_{HD} = L_{BB}/L_{AD}$, i.e. the ratio at $\lambda$ = 2 \textmu m between the hot-dust and accretion disk components.\\
In detail, we modeled the QSOs SEDs at $\lambda > 3000$ \AA\ with a power-law describing the emission from the accretion disk plus a single temperature black-body which represents the emission from hot dust located in the innermost layer of the torus, i.e.:
\begin{equation*}
\lambda L_{\lambda} = K_{1}\lambda^{\gamma_\text{opt}} + K_{2}\frac{1}{\lambda^{4}}\frac{1}{\exp(hc/\lambda k_{B}\text{T})-1}  
\end{equation*}
Following \cite{Temple:2021}, we fixed both the temperature of the Black-body (T = 1280 K) and the slope of the power-law ($\gamma_\text{opt} = -0.16$). We kept T fixed since the only point useful to constrain the black-body emission is the Spitzer 24 \textmu m one (or, alternatively, W4), while we fixed $\gamma_\text{opt}$ to compare our results against those by \cite{Temple:2021} who showed that derived $X_{HD}$ values strongly depends on the assumed $\gamma_\text{opt}$. However, we note that the assumed slope is remarkably close to the one we obtain in Sect. \ref{sec:mean_sed} for the mean SED and therefore can be considered as an average value, although it is slightly different from the mean of the individual $\gamma_\text{opt}$ computed in Sect. \ref{sec:slopes} which is $\langle \gamma_\text{opt} \rangle = -0.23$; in addition, by visually inspecting the fitting results we found that the optical emission of the analyzed 11 sources is generally well described by this power-law.
As done for the SED fittig routine, the constants of normalization $K_{1}$ and $K_{2}$ were derived by minimizing the likelihood.
\begin{figure*}[h!]
    \sidecaption
    \includegraphics[width = 12cm]{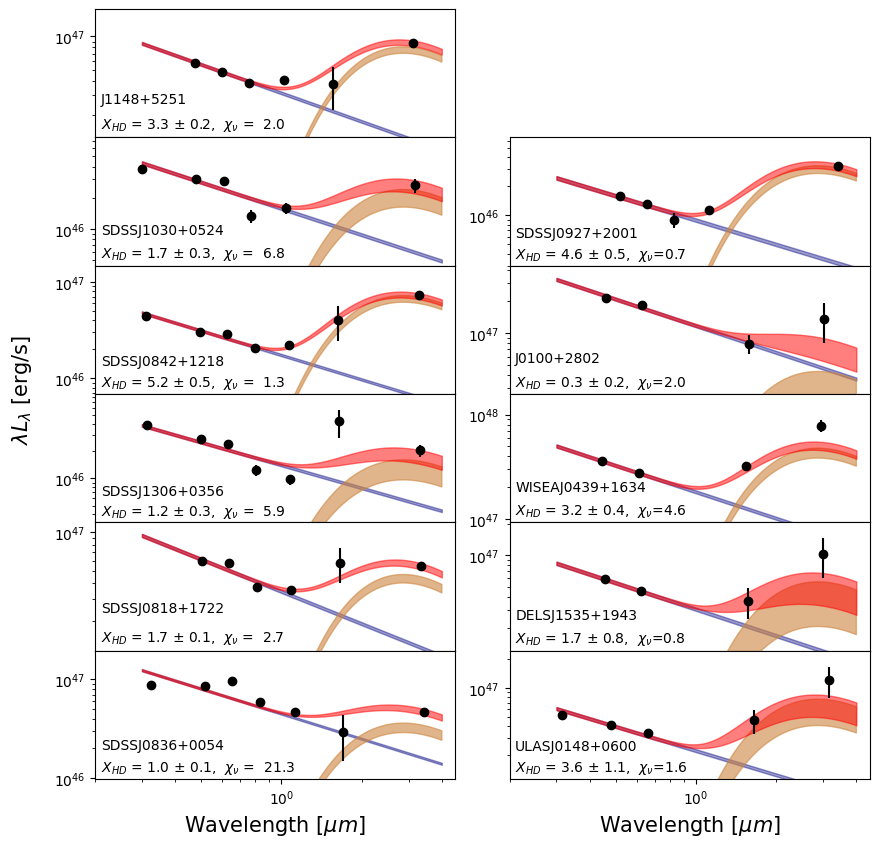}
    \caption{NIR SEDs for the 11 QSOs with MIR photometry. The blue line gives the accretion disk emission modeled as a power-law with a spectral  index of -0.16 while the gold line describes emission from hot dust modeled as a blackbody with T = 1280 K. The red lines show the sum of the two components.}
    \label{fig:nir_seds}
\end{figure*}
Results of the fitting are shown in Fig. \ref{fig:nir_seds}. 
For each of the analyzed QSOs, it is necessary to add a hot dust component, and therefore there are no QSOs free of dust component in our sample. We find the median value of $X_{HD}$ to be 2.4 with a median absolute deviation of 1.4  which is in agreement with the 2.5 median value reported in \cite{Temple:2021}; moreover, a K-S test (p-value = 0.19) indicates no evidence for a different underlying $X_{HD}$ distribution compared to that derived by \cite{Temple:2021}.\\
Specifically, two sources, namely J0100+2802 and SDSSJ0836+0054, exhibit $X_{HD} \leq 1$. While only J0100+2802 can be considered, within 1$\sigma$, as properly dust-poor according to the criterion set by \cite{Jun:2013}, which requires $X_{HD} \leq 0.15$ \citep[see][]{Temple:2021}, both sources clearly show much lower dust emission than expected. Notably, SDSSJ0836+005 was previously identified as a QSO with dust deficiency by \cite{Leipski:2014}.
On the other hand, three QSOs, ULASJ0148+0600, SDSSJ0842+1218 and SDSSJ0927+20, are found to have $X_{HD} \geq 4.7$.\\
\begin{figure}
    \centering
    \includegraphics[width = 0.48\textwidth]{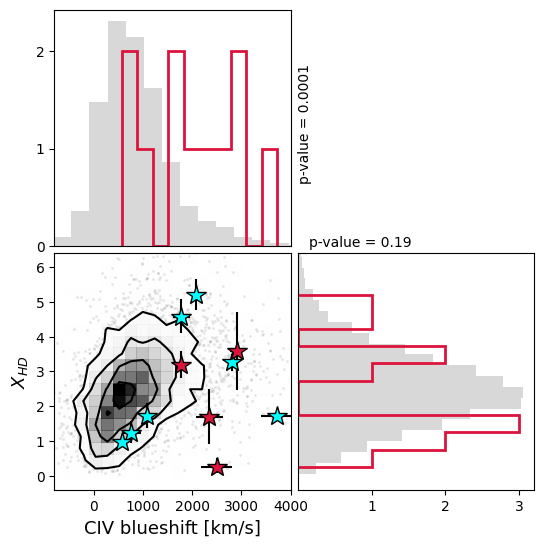}
    \caption{$X_{HD}$ vs CIV blueshift, adapted from Fig. 3 in \cite{Temple:2021} with the inclusion of HYPERION and XQR-30 QSOs. E-XQR-30 blueshifts are taken from \cite{Mazzucchelli:2023}, while HYPERION values are computed as the mean of all the blueshifts reported in the literature for each QSO, see \cite{Tortosa:2024b}. Errors on $X_{HD}$ are given as the 16th and 84th percentiles. Cyan stars indicate QSOs with Spitzer MIPS 24 \textmu m fluxes while red stars have W4 photometry. The histograms show the  distributions of CIV blueshift and $X_{HD}$ in \cite{Temple:2021} (gray, normalized by a factor of 500) and for the HYPERION and E-XQR-30 QSOs (red). The p-values obtained by performing a K-S test on the two distributions are reported next to the histograms.}
    \label{fig:civ_nir}
\end{figure}
\cite{Temple:2021} reported a positive correlation between $X_{HD}$  and the velocity of the CIV emission line relative to the systemic MgII-derived redshift (hereafter, CIV blueshifts), i.e. objects with faster winds have stronger hot-dust emission; Fig. \ref{fig:civ_nir} shows the addition of the 11 HYPERION and E-XQR-30 sources to their original plot. Our sources do not show any clear CIV-$X_{HD}$ correlation. However this result is primarily due to two factors. Firstly, the very small sample size poses challenges in determining the correlation. As a test, we randomly selected 11 sources from the data presented in \cite{Temple:2021}, and only in 25\% of the cases we recovered a positive correlation with a p-value $<$ 0.05. Secondly, more importantly, our sources exhibit significantly faster winds compared to the QSOs in \cite{Temple:2021}, as expected for such luminous sources \citep[][]{Fiore:2017, Vietri:2018, Meyer:2019, Timlin:2020, Schindler:2020}. Consequently, a majority of our objects lie outside the contour enclosing 86.4\% (i.e. 2$\sigma$) of \cite{Temple:2021} QSOs.

\section{The mean spectral energy distribution of luminous, z$>$6 QSOs}
\label{sec:mean_sed}
We generated a mean SED based on the combined HYPERION and E-XQR-30 samples to obtain a more comprehensive view of the broadband emission of luminous high-redshift QSOs. Since our sample consists of QSOs with a relatively narrow redshift range, each observed band probes similar rest-frame wavelengths across all QSOs. Therefore, instead of constructing a continuous SED template, we determine the mean luminosity value for each band. With this approach, we do not need to extrapolate the SED to wavelengths not covered by any QSOs. We also grouped as the same band  Spitzer IRAC 3.6 and W1, Spitzer IRAC 4.5 and W2, Spitzer MIPS 24 and W4; J1120+0641 1 \textmu m luminosity point was included with Spitzer MIPS 8 \textmu m.
To minimize the scatter due to the small sample size of our analyzed QSOs, we have normalized their SEDs at 3500 \AA. The normalization is particularly relevant for the W3 and W4 bands, where most sources in our sample have not been detected. Indeed, since brighter QSOs are more likely to be detected, using non-normalized data would result in a mean NIR emission that is not representative of the true mean properties of the sources.
Although there is no physical reason to choose a specific normalization wavelength, we tested various wavelengths within the range of 2000 \AA < $\lambda$ < 7000 \AA\ and found that normalizing at 3500 \AA\ resulted in the lowest scatter across all bands. Therefore, we adopted this wavelength as our normalization point.
\begin{figure}
    \centering
    \includegraphics[width =0.49\textwidth]{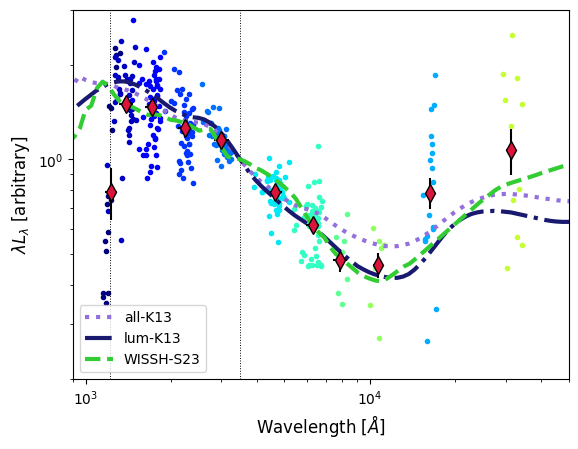}
    \caption{Photometry points for the HYPERION and E-XQR-30 samples, normalized at $\lambda$ = 3500 \AA. The red diamonds show the geometric mean value for each band with the associated 1$\sigma$ as error bar. Overplotted are also the mean SEDs by \cite{Krawczyk:2013} (All and luminous samples, purple dotted and blue dot-dashed lines respectively) and the WISSH hyper-luminous mean SED (dashed green line). The vertical dotted lines indicate the position of the Ly$\alpha$ at 1216 \AA\ and the normalization wavelength.}
    \label{fig:Mean_SED}
\end{figure}

\begin{figure}
    \centering    
    \includegraphics[width = 0.48\textwidth]{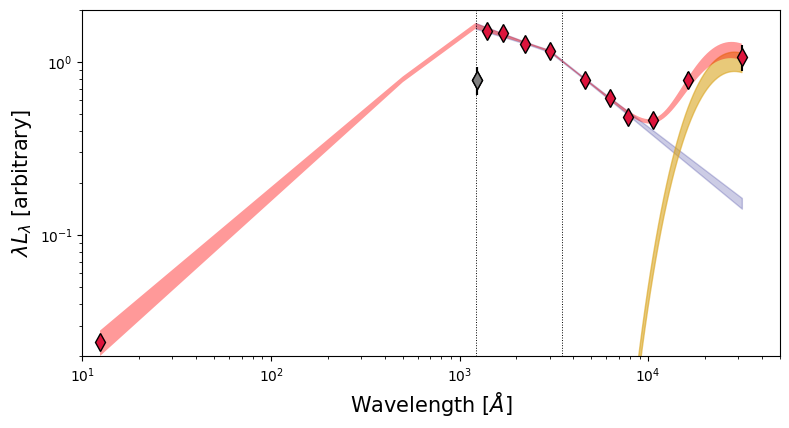}
    \caption{Continuous mean SED obtained by fitting the mean points (red diamonds) with a broken power-law ($\lambda_{break}$ = 3000 \AA) plus a single temperature (T = 1280 K) black-body. The EUV SED ($\lambda < 1216$ \AA) was reconstructed as discussed in Sect. \ref{sec:sed_fitting}. The light blue and gold shaded area describe the best-fit broken power-law and black-body respectively. The grey diamond was excluded from the fit since it is the average also of points falling below the Ly$\alpha$.
    The black dotted lines indicate the Ly$\alpha$ and the normalization wavelength at 3500 \AA.}
    \label{fig:mean_SED_completa}
\end{figure}

Fig. \ref{fig:Mean_SED} shows the resulting mean SED, with the uncertainties given as the standard deviation divided by the square root of the number of QSOs in each band, along with a comparison with the mean SEDs derived in \cite{Krawczyk:2013} for both their luminous and whole QSO sample (dot-dashed blue line and dotted purple line respectively) and WISSH-S23 (dashed green line). \\
 As visible in Fig. \ref{fig:Mean_SED}, we do not find any significant deviations from the mean SEDs of luminous QSOs at lower-\textit{z} (i.e. lum-K13 and WISSH-S23) while the average SED by \cite{Krawczyk:2013}, labeled as all-K13,  has a flatter UV-optical slope and, depending on the chosen normalization wavelength, predicts larger emission at $\lambda \sim 6000-9000$ \AA\, as in Fig. \ref{fig:Mean_SED}, or a lower UV bump. Therefore we do not find any evolution of the SED with redshift, at least for the 1000 \AA\ - 1 \textmu m  wavelength interval. 
While there might be a selection effect, given that all our sources were identified by targeting objects with typical colors of lower-\textit{z} QSOs \citep[e.g.][]{Reed:2019}, and therefore may not be universally applicable to the entire population of high-z AGN, this result strengthens our choice to employ templates of luminous QSOs for computing bolometric luminosities.\\
At wavelengths above 1 \textmu m  the distribution of rest-frame NIR points shows a considerable spread; there is an indication for a NIR emission even stronger than in the IR selected WISSH QSOs which needs to be further investigated, as the current limited data prevents us from doing any claims.
Moreover, the W3 and W4 points are likely biased towards NIR-bright sources, as the limiting magnitudes in these bands are much shallower than those of W1 and W2. \\
We fitted the mean SED with a broken power-law jointed at $\lambda$ = 3000 \AA\ plus a blackbody with a fixed temperature T = 1280 K as described in Sect. \ref{sec:NIR}. For all parameters we assumed flat priors. 
We find $\beta_\text{UV} = -0.63 \pm 0.05$ which is in agreement with both the value reported in \cite{Telfer:2002} (i.e -0.69, from the analysis of 184 QSOs with HST spectra) and the one presented by \cite{Lusso:2015} (i.e. -0.61, derived from 53 luminous sources at redshift $z \sim 2.4$). Such a value of $\beta_\text{UV}$ is instead softer than what is found in the composite spectrum by \cite{Selsing:2016} derived from 102  QSOs at \textit{z} = 1-2.1 and by \cite{Temple:2021b} from a subsample of bright SDSS QSOs (18.6 $<$ i $<$ 19.1), i.e. $\beta_\text{UV} =$ -0.30 and -0.349 respectively.
The power-law modeling the mean SED redwards of 3000 \AA\ is found to be  steeper with a slope $\gamma_\text{opt} = -0.14 \pm 0.04$. 
Interpreting $\gamma$ as the mean optical slope we confirm that high-luminosity QSOs have a steeper optical continuum \citep[e.g.][]{Richards:2006, Krawczyk:2013} even at high redshift with respect to the bulk of the population at lower luminosities \citep[i.e. $\gamma_\text{opt}$ = -0.46][]{Vanden-Berk:2001}. 
Interestingly, computing the hot dust-to-AD ratio with this modeling gives $X_{HD} = 3.28 \pm 0.5$, which is higher than both the median value found by \cite{Temple:2021} and that calculated as the mean of the individual QSOs. \\
The continuous SED template 
obtained by the broken power-law plus black-body modeling is shown in Fig. \ref{fig:mean_SED_completa}. To connect the template with the mean 1 kev luminosity we extend the SED in the EUV region by using the same recipe reported in eq. \ref{eqn:double_PL}.\\

\begin{figure}
    \centering    
    \includegraphics[width = 0.48\textwidth]{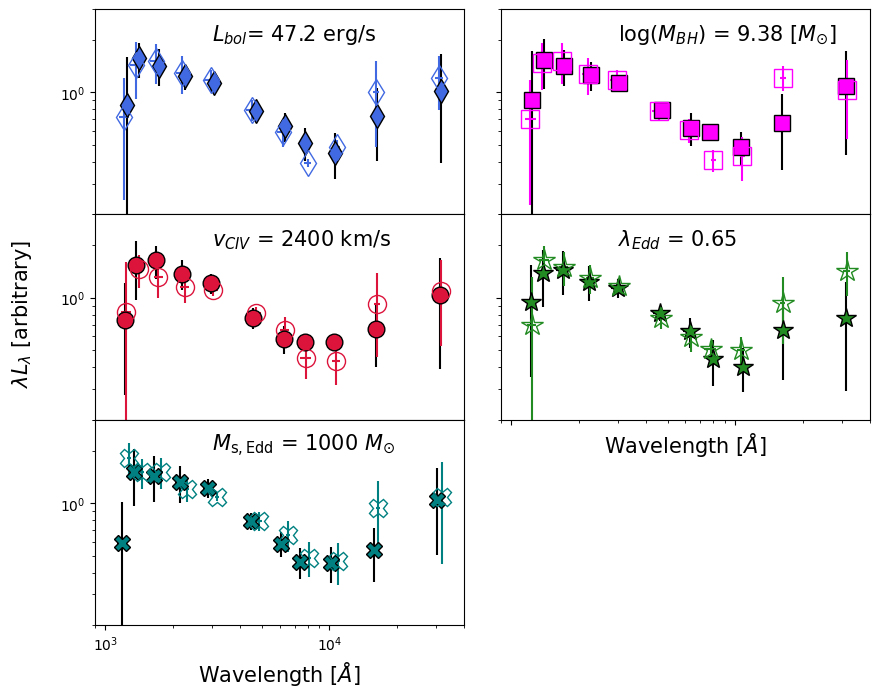}
    \caption{Mean SEDs derived by splitting the sample according to the median value of several physical properties. Each panel reports the threshold value used to split the sample. Filled markers refer to QSOs above the threshold while empty ones indicate QSOs below the threshold.}
    \label{fig:mean_SED_parameters}
\end{figure}

\begin{figure}
    
    \centering
    \includegraphics[width = 0.48\textwidth]{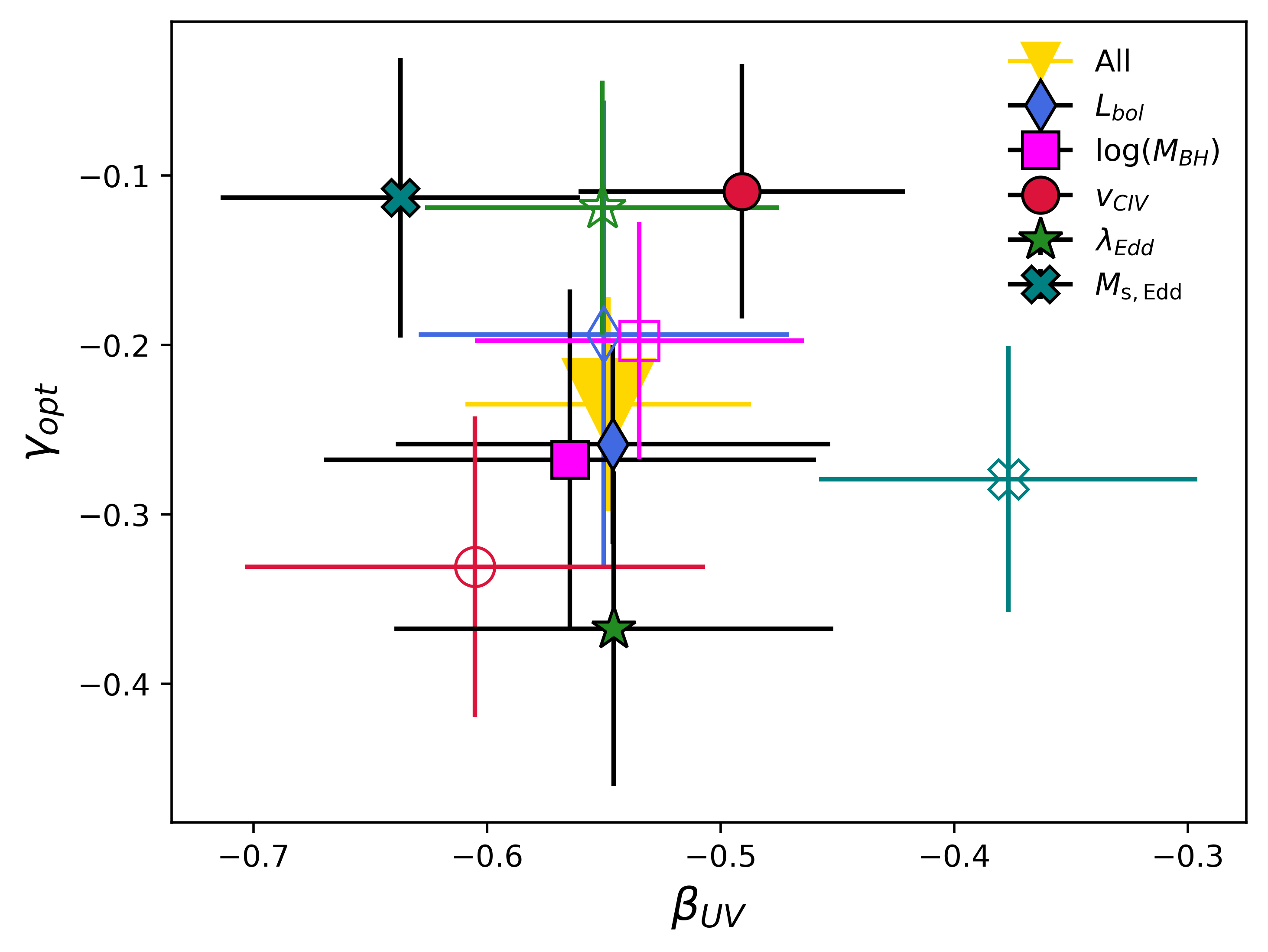}
    \caption{Mean $\gamma_\text{opt}$ vs mean $\beta_\text{UV}$ computed for each subsample. Errors are reported as the standard deviation of the values divided by the square root of the number of sources in the subsample. The yellow triangle shows the mean value for the full sample. As in Fig. \ref{fig:mean_SED_parameters} filled markers refer the subsample above the threshold while empty ones refer to the one below.}
    \label{fig:sed_slope_plot}
\end{figure}

We also computed the  mean SEDs of several subsamples obtained by splitting the sample based on the median of several physical properties ($M_\text{SMBH}$, CIV blueshift, $L_\text{bol}$) or on the criterion adopted to assemble the HYPERION sample (i.e. using $M_{\rm s, Edd}$=1000 $M_{\odot}$ as threshold) but found no significant difference either between them or with respect to the overall average SED, (see Fig. \ref{fig:mean_SED_parameters}). This result holds true even when comparing the mean values of $\beta_\text{UV}$ and $\gamma_\text{opt}$ of each subsample, as shown in Fig. \ref{fig:sed_slope_plot}. Indeed , each pair of values is in agreement with that of its complementary subsample within 2$\sigma$.

\section{Summary and Conclusions}
In this work we characterize the  X-ray-to-NIR broad-band emission of luminous QSOs at the EoR with the main goal of providing the first systematic investigation of the SED of these sources.
Our main results are here summarized:
\begin{itemize}
\item We produced X-ray-to-NIR  SEDs for the 18 QSOs belonging to the HYPERION sample \citep[][]{Zappacosta:2023, Tortosa:2024b}. Exploiting the unprecedented quality of X-ray data from the HYPERION QSOs, we employed a double power-law to characterize the X-ray to UV SED \citep[][see eq. \ref{eqn:double_PL}]{Lusso:2012};  given the absence of data in that wavelength range, this modeling  ensures the most reliable representation of the EUV region that we can achieve.
The UV-to-NIR SED of the QSOs was instead modeled using templates derived from luminous QSOs at lower-\textit{z}, i.e. lum-K13 \citep[][]{Krawczyk:2013} and WISSH-S23 \citep[][]{Saccheo:2023} We found that for all QSOs the optical-UV region is well described by the lum-K13 SED (see Fig. \ref{fig:fitting}) while we obtained, on average, worse result when employing WISSH-S23 SED although they span a luminosity range closer to the HYPERION sources.
\item We increase the statistical significance of our analysis in the UV-to-NIR region by including additional 36 QSOs drawn from the E-XQR-30 sample \citep[][]{DOdorico:2023} which share similar properties with HYPERION sources in terms of redshift and luminosities. We confirm that also for these sources lum-K13 template provides excellent results in describing their SED.

\item By integrating the SEDs we computed bolometric luminosities homogeneously for all 54 objects. Their values, shown in Fig. \ref{fig:luminosities_histograms} range from 46.8 $\leq log(L_\text{bol}) \leq$ 48.1. We found that our derived $L_\text{bol}$ are systematically lower than the ones already reported in the literature, with an average discrepancy factor of 0.13 dex.

\item We calculated bolometric corrections at 3000, 4400, and 5100 \AA\ for QSOs with redshifts z$\geq 5.5$ and check their consistency with those derived from lower-\textit{z} sources. Given the similarity of the SEDs of our QSOs to those of their lower-\textit{z} counterparts, the obtained bolometric corrections are broadly consistent with literature findings. However, we observe that the values for $\lambda$ = 3000 \AA\ and 5100 \AA\ are systematically 
larger than the bolometric correction values largely used in the literature. As an improved estimate for these sources, we showed that better results are obtained by using $BC_{3000\text{\AA}}$ = 3.3 and $BC_{5100\text{\AA}}$ = 5.7 to derive $L_\text{bol}$.

\item Taking advantage of HYPERION  X-ray coverage we verified the accuracy of theoretical models \citep[][]{Kubota:2018} in predicting coronal emission from that of the accretion disk for these high-\textit{z} sources. Despite the many crude assumptions we made on the parameters, we surprisingly find that models and observations deviate by a median factor of just 1.6.

\item We investigated the hot dust emission in 11 QSOs with available MIR data by modeling their rest-frame NIR  using a combination of a power-law and a blackbody (see Sect. \ref{sec:NIR} and Fig. \ref{fig:nir_seds}). Our analysis revealed that, for all these sources, the inclusion of a dust component is necessary, although with varying strengths. We quantified the hot dust emission strength by computing the hot dust to accretion disk emission ratio $X_{HD}$ \citep[][]{Temple:2021}. The overall $X_{HD}$ distribution is consistent with the one found by \cite{Temple:2021}, see Fig. \ref{fig:civ_nir}, but two QSOs exhibit a notably low $X_{HD}$, with one source, J0100+2802, being within 1$\sigma$ of the threshold outlined in \cite{Jun:2013} for classification as dust-poor.
On the other hand, 3 sources have $X_{HD}$ about twice the average one observed in luminous QSOs.

\item Finally, we have derived a mean SED for these high-\textit{z} QSOs extending from the X-ray to NIR (i.e. from 1 keV to about 3 \textmu m), see Fig.\ref{fig:Mean_SED} and \ref{fig:mean_SED_completa}. This average SED is in excellent agreement with the mean SEDs of luminous QSOs, pointing out that for the analyzed sample  we do not find any redshift evolution of the SED's shape, at least in the optical-UV region. As typically reported for luminous QSOs at lower-\textit{z}, we found that optical emission at 3000 \AA\ $< \lambda <$ 1 \textmu m is well described by a power-law with a steeper slope ($F_{\nu} \propto \nu^{-0.14}$) than observed in the bulk of low-luminosity AGN population, i.e. $F_{\nu} \propto \nu^{-0.46}$, \citep[][]{Vanden-Berk:2001}.  
\end{itemize}
As a natural extension of this work, we aim to (i) increase the sample size of QSOs at z$>$ 6-7 with good X-ray data and (ii) include rest-frame photometric points at $\lambda>$1 \textmu m thanks to JWST/MIRI observations in  order to have an uniform NIR coverage.
Improving these aspects will be fundamental for highlighting potential redshift-dependent properties, which we have so far been unable to constrain in a significant way. To this end, a recently accepted XMM-Newton Large Program (PI Zappacosta) will allow the investigation of X-ray emission in also in sources with low $M_{\rm s, Edd}$. In the NIR, instead, it will be critical to increase the number of sources with photometric coverage to confirm the tentative indication from the computed mean SED that, on average, EoR sources exhibit enhanced emission from hot dust compared to average templates. JWST/MIRI observations will play a crucial role in adequately sampling the SED in the 1 to 2.5 \textmu m region, providing a systematic characterization of hot dust emission in these early quasars.

\begin{acknowledgements}
We are grateful to the anonymous referee for their useful comments and suggestions which helped us to improve the paper.
The authors acknowledge financial support from the Bando Ricerca Fondamentale INAF 2022 Large Grant "Toward an holistic view of the Titans: multi-band observations of z $>$ 6 QSOs powered by greedy supermassive black holes". 
AB, MB, MB, SC, VD, FF, CF, SG, VT, NM, LZ acknowledge support from
the European Union - Next Generation EU, PRIN/MUR  2022 2022TKPB2P - BIG-z.
MB acknowledges support from INAF project 1.05.12.04.01 - MINI-GRANTS di RSN1 "Mini-feedback" and from UniTs under FVG LR 2/2011 project D55-microgrants23 "Hyper-gal".
DD acknowledges PON R\&I 2021, CUP E65F21002880003. FT acknowledges funding from the European Union - Next Generation EU, PRIN/MUR 2022 2022K9N5B4. 
GM acknowledges financial support by grant PID2020-115325GB-C31 funded by MICIN/AEI/10.13039/501100011033. This work was supported by STFC grant ST/X001075/1
\end{acknowledgements}

\bibliographystyle{aa}
\bibliography{ivano}

\newpage
\appendix
\section{Proprietary NIR observations and Data reduction}

\label{sec:proprietary_observations}

\begin{table}
    \centering
\setlength{\tabcolsep}{3pt} 
\renewcommand{\arraystretch}{1.1} 
\scalebox{0.90}{
\begin{tabular}{lcllc}
\textbf{Name}	& \textbf{Instrument}	 & \textbf{Filters}	& \textbf{Exp. time} & \textbf{Date}  \\
\toprule
J1342+0928                & NICS   &	  J, H, K      &   62, 60,  123    &   25/05  \\ \hline
J1120+0641	                & NICS	 &	  J,H,K        &   7, 15, 10       &   18/04  \\ \hline
\multirow{2}{*}{J0038-1527}	& SOFI	 &    J, H, K	   &   12, 6, 6	       &   03/11  \\
	                              & NICS   &    J, H, K	     &   6, 7, 9	     &   02/11  \\ \hline
\multirow{2}{*}{J0252-0503}	& SOFI	 &    J, H, K	   &   7, 10, 15	   &   03/11  \\
                                & NICS	 &    Y, J, H, K   &   10, 15, 18, 22  &   02/11  \\ \hline
J0020-36	                & SOFI	 &    J, H, K      &   9, 15, 15       &   03/11  \\ \hline
\multirow{2}{*}{J0411-0907}  & SOFI	 &    J, H, K	   &   18, 21, 30	   &   03/11  \\ 
                                & NICS	 &    Y, J, H, K   &   6, 11, 11, 12   &   02/11  \\ \hline
J0244-5008                & SOFI	 &    J, H, K	   &   21, 10. 10	   &   03/11  \\ \hline
J231-20.8               & NICS   &    Y, J         &   10, 23          &   24/04  \\ \hline
J036+03.0                & NICS	 &    Y, J, H, K   &   5, 15, 20, 31   &   16/09  \\ \hline
J0224-4711                & SOFI	 &    J, H, K	   &   6, 6, 6	       &   03/11  \\ \hline
\multirow{3}{*}{J011+09}     & NICS   &    Y, J, H      &   44, 50, 40      &   04/09  \\
                                & NICS	 &    K	           &   60	           &   16/09  \\ 
	                              & NICS   &    Y, J, H, K   &   21, 44, 50, 60	 &   02/11 \\  \hline 
J083+11.8                & SOFI	 &    J, H, K	   &   18, 21, 27	   &   03/11  \\ \hline
\multirow{2}{*}{J0050+3445}    	& NICS 	 &    Y, J, H, K   &   15, 20, 30, 30  &   16/09  \\
	                            & NICS	 &    Y, J, H, K   &   6, 10, 11, 16   &   02/11  \\ \hline
J029-36  	                & SOFI	 &    J, H, K	   &   18, 18, 18	   &   03/11 \\ 
\bottomrule

\end{tabular}
}
    \caption{Diary of NIR imaging observations for the HYPERION sample. Exposure times are in minutes. All observations were conducted in 2022.}
    \label{tab:observations}
\end{table}

New NIR (Y, J, H and K bands) observations have been obtained for 15 sources with NICS at the INAF funded telescope TNG during programmes AOT45 and AOT46 (PI L. Zappacosta and F. La Franca respectively) and with SOFI at ESO/NTT, run 110.244M.001, (PI I. Saccheo). Details about the observations and the net exposure time per band are reported in Tab. \ref{tab:observations}.
Data were collected 
adopting a dithering strategy and acquiring at least 5 images per filter and were reduced with the following steps:
we first applied a median stacking technique with a clipping algorithm to produce an empty sky-frame. This frame was then subtracted from all the raw frames to obtain a set of images with a median background level equal to zero. This step also removes bias and dark current levels. We then divided the sky-subtracted frames by the normalized flat-field image, obtained by twilight sky frames
(for TNG) or already provided by the telescope staff (SpecialDomeFlat, see SOFI User manual). The flat-fielded images were then astrometrically registered and co-added obtaining the final frames for each filter. We performed photometry extraction using the python package \textit{photutils} \citep[][]{Bradley:2022} and utilized an 8 pixel radius (e.g. $\sim$ 2 arcsec) to perform aperture photometry since we deal with point-like sources in not crowded fields. Errors on fluxes were computed as is done in IRAF package  DAOPHOT, i.e. $\sigma^{2}_{f} = f/epadu +A_{f}\sigma^{2}_{bkg}( 1+ A_{f}/A_{bkg})$ where \textit{f} is the flux measured in the aperture area $A_{f}$, $\sigma_{bkg}$ is the standard deviation of the background computed in a circular annulus of area $A_{bkg}$ and \textit{epadu} is the telescope gain expressed in electrons per count.
Instrument magnitudes were then calibrated by using nearby standard stars observed immediately before and after the QSO observations (J, H and K bands for NTT and Y band at TNG) or by using stars in the field also observed in the 2MASS survey \citep[J, H and K bands at TNG][]{Cutri:2003}. For sources observed at both TNG and NTT, we found that SOFI images
provided more accurate (in terms of S/N ratio) but still compatible data and thus we report the SOFI measurements.
In 3 cases (J1120+06, J231-20 and J036+03) observations were significantly  affected by poor sky conditions (i.e. seeing > 1.7 arcsec); we discarded those observations.\\

\section{HYPERION magnitudes}
We report the NIR-to-MIR AB photometry from proprietary observations and taken from the literature for HYPERION QSOs.
Tab. \ref{table:photometry_master} presents the data in band z, Y, J, H, K, W1, W2, W3 and W4.
If available,  the photometric data from Spitzer in the IRAC bands (3.6, 4.5 and 5.8 \textmu m) and  in the MIPS band at 24 \textmu m are also listed.

\begin{landscape}
\begin{table}
\addtolength{\tabcolsep}{-3pt}  
\renewcommand\arraystretch{1.3}
\caption{Near and Mid IR AB magnitudes of the HYPERION QSOs.}
\begin{tabular}{lcccccccccc}

\toprule

\multirow{2}{*}{Name} &     \multirow{2}{*}{z}   &   \multirow{2}{*}{Y}      &  \multirow{2}{*}{J}     &      \multirow{2}{*}{H}          &       \multirow{2}{*}{K}     & W1$^{\star}$ & W2$^{\star}$ & W3$^{\ast}$ & W4$^{\ast}$ & \multirow{2}{*}{S$_{24\mu m}$} \\
& & & & & & S$_{3.6\mu m}$ & S$_{4.5\mu m}$ & S$_{5.8\mu m}$ & S$_{8.0\mu m}$ & \\

\midrule
J1342+0928 & $>$  23.32$^{1}$        & 21.44 $\pm$ 0.19 $^{2}$      & 20.06 $\pm$ 0.05 $^{\dag}$ & 20.04 $\pm$ 0.05 $^{\dag}$ & 19.97 $\pm$ 0.05 $^{\dag}$ & 20.32 $\pm$   0.08 & 20.32 $\pm$   0.18 &       &       & \\ 
J1007+2115     & $>$23.02$^{3}$          & 21.30 $\pm$ 0.13 $^{3}$      & 20.2 $\pm$  0.18 $^{3}$    & 20.00 $\pm$ 0.07 $^{3}$    & 19.75 $\pm$ 0.08 $^{3}$    & 19.71 $\pm$   0.05 & 19.65 $\pm$   0.11 &       &       & \\
\hline
\multirow{2}{*}{J1120+0641} & \multirow{2}{*}{23.19 $\pm$ 0.06$^{4}$}  & \multirow{2}{*}{20.27 $\pm$ 0.09 $^{2}$}      & \multirow{2}{*}{20.27 $\pm$ 0.15 $^{2}$}    & \multirow{2}{*}{19.60 $\pm$ 0.14 $^{2}$} & \multirow{2}{*}{19.54 $\pm$ 0.17 $^{2}$} & 19.64 $\pm$ 0.05 & 19.81 $\pm$ 0.14  &       &        &\\
& & & & & & 19.39 $\pm$ 0.03 $^{4}$& 19.49 $\pm$ 0.03$^{4}$ & & 19.48 $\pm$ 0.06 $^{19}$ & \\
\hline
J0038-1527 & 21.65 $\pm$ 0.08$^{5}$  & 20.02 $\pm$ 0.08 $^{5}$      & 19.66 $\pm$ 0.05 $^{\dag}$ & 19.51 $\pm$ 0.09 $^{\dag}$ & 19.28 $\pm$ 0.11 $^{\dag}$ & 19.44 $\pm$   0.04 & 19.65 $\pm$   0.11 &       &       &\\
J0252-0503 & 22.34 $\pm$ 0.09$^{6}$   & 20.33 $\pm$ 0.07 $^{7}$      & 20.18 $\pm$ 0.08 $^{\dag}$ & 19.91 $\pm$ 0.1  $^{\dag}$ & 19.95 $\pm$ 0.11 $^{\dag}$ & 20.11 $\pm$   0.08 & 20.24 $\pm$   0.19 &       &       &\\
J0020-3653 & 21.46 $\pm$ 0.05$^{6}$  & 20.16 $\pm$ 0.05 $^{6}$      & 20.36 $\pm$ 0.10 $^{8}$    &                            & 19.55 $\pm$ 0.13 $^{8}$    & 19.57 $\pm$   0.04 & 19.75 $\pm$   0.11 &       &      & \\
J0411-0907  & $>$ 22.3$^{9}$          & 19.94 $\pm$ 0.10 $^{\dag}$   & 19.72 $\pm$ 0.04 $^{\dag}$ & 19.51 $\pm$ 0.06 $^{\dag}$ & 19.47 $\pm$ 0.07 $^{\dag}$ & 19.52 $\pm$   0.04 & 19.39 $\pm$   0.08 &       &       & \\
J0244-5008 & 21.17 $\pm$ 0.03$^{6}$  & 20.14 $\pm$ 0.04 $^{6}$      & 20.23 $\pm$ 0.07 $^{\dag}$ & 19.99 $\pm$ 0.09 $^{\dag}$ & 19.78 $\pm$ 0.07 $^{\dag}$ & 19.66 $\pm$   0.04 & 19.51 $\pm$   0.07 &       &       & \\
J231-20.8 & $>$ 22.77$^{10}$        & 20.14 $\pm$ 0.08 $^{10}$     & 19.66 $\pm$ 0.05 $^{10}$   &          19.55 $\pm$ 0.01 $^{11}$                  & 19.52 $\pm$ 0.18 $^{2}$    & 19.91 $\pm$   0.15 & 19.96 $\pm$   0.35 &       &      & \\
J036+03.0 & 20.01 $\pm$ 0.01$^{10}$ & 19.39 $\pm$ 0.04 $^{6}$      & 19.51 $\pm$ 0.03 $^{12}$   & 19.29 $\pm$ 0.09 $^{2}$    & 19.23 $\pm$ 0.09 $^{2}$    & 19.36 $\pm$   0.04 & 19.52 $\pm$    0.1 &       &       & \\
J0224-4711 & 20.20 $\pm$ 0.02$^{6}$  & 20.03 $\pm$ 0.04 $^{6}$      & 19.80 $\pm$ 0.08 $^{\dag}$ & 19.43 $\pm$ 0.11 $^{\dag}$ & 18.85 $\pm$ 0.09 $^{\dag}$ & 18.81 $\pm$   0.02 & 18.76 $\pm$   0.04 &       &       &  \\
J011+09     & 22.38 $\pm$ 0.16$^{10}$ & 20.51 $\pm$ 0.04 $^{\dag}$   & 20.36 $\pm$ 0.03 $^{\dag}$ & 20.55 $\pm$ 0.05 $^{\dag}$ & 19.94 $\pm$ 0.06 $^{\dag}$ & 20.47 $\pm$   0.11 &        &     &   & \\
\hline
\multirow{2}{*}{J1148+5251} & \multirow{2}{*}{20.56 $\pm$ 0.03$^{13}$} & \multirow{2}{*}{19.01 $\pm$ 0.04 $^{13}$}     & \multirow{2}{*}{19.08 $\pm$ 0.06 $^{14}$}   & \multirow{2}{*}{19.01 $\pm$ 0.05 $^{15}$}   & \multirow{2}{*}{18.83 $\pm$ 0.05 $^{15}$}  & 18.76 $\pm$   0.02 & 18.77 $\pm$   0.04 & 17.73 $\pm$   0.45 & 15.53 $\pm$   0.47 &   \multirow{2}{*}{16.07 $\pm$ 0.04$^{20}$}\\
& & & & & & 18.57 $\pm$ 0.02$^{20}$ & 18.51 $\pm$ 0.02$^{20}$ & 18.50 $\pm$ 0.05 $^{20}$ & 18.10 $\pm$ 0.04$^{20}$ &  \\
\hline
J083+11.8 & 21.72 $\pm$ 0.17$^{16}$ & 19.91 $\pm$ 0.06 $^{16}$     & 19.70 $\pm$ 0.07 $^{\dag}$ & 19.45 $\pm$ 0.07 $^{\dag}$ & 19.42 $\pm$ 0.09 $^{\dag}$ & 19.37 $\pm$   0.04 & 19.61 $\pm$   0.11 &    &    &  \\ 
J0100+2802     & 18.61 $\pm$ 0.01$^{13}$ & 17.62 $\pm$ 0.01 $^{13}$     & 17.60 $\pm$ 0.02 $^{17}$   & 17.48 $\pm$ 0.02 $^{17}$   & 17.05 $\pm$ 0.16 $^{18}$   & 17.14 $\pm$   0.01 & 16.98 $\pm$   0.01 & 16.89 $\pm$   0.21 &  15.6 $\pm$   0.44 &\\
J025-33   & 19.52 $\pm$ 0.01$^{17}$ & 19.14 $\pm$ 0.01 $^{17}$     & 19.11 $\pm$ 0.01 $^{17}$   & 18.95 $\pm$ 0.02 $^{17}$   & 18.75 $\pm$ 0.03 $^{17}$   & 18.83 $\pm$   0.02 &  18.7 $\pm$   0.04 & 17.76 $\pm$   0.43 &    &  \\
J0050+3445     & 20.98 $\pm$ 0.07$^{15}$ & 19.98 $\pm$ 0.05 $^{\dag}$   & 19.90 $\pm$ 0.08 $^{\dag}$ & 19.35 $\pm$ 0.07 $^{\dag}$ & 19.21 $\pm$ 0.09 $^{\dag}$ & 19.28 $\pm$   0.03 & 19.05 $\pm$   0.06 &         &      & \\ 
J029-36   & 19.65 $\pm$ 0.01$^{6}$  & 19.79 $\pm$ 0.04 $^{6}$      & 19.47 $\pm$ 0.04 $^{\dag}$ & 19.59 $\pm$ 0.08 $^{\dag}$ & 19.02 $\pm$ 0.08 $^{\dag}$ & 19.31 $\pm$   0.03 & 19.25 $\pm$   0.07 &         &       &  \\    \bottomrule
\end{tabular}
\addtolength{\tabcolsep}{-1pt} 
\tablefoot{References:1 -\cite{Banados:2018}, 2-\cite{Lawrence:2012}, 3-\cite{Yang:2020}, 4-\cite{Barnett:2015}, 5-\cite{Wang:2018}, 6-\cite{Abbott:2021}, 7-\cite{Wang:2020}, 8-\cite{McMahon:2021}, 9-\cite{Pons:2019}, 10-\cite{Mazzucchelli:2017},
11-\cite{DOdorico:2023},
12-\cite{Venemans:2015}, 13-\cite{Banados:2016}, 14-\cite{Shen:2019}, 15-\cite{Iwamuro:2004}, 16-\cite{Andika:2020}, 17-\cite{Ross:2020}, 18-\cite{Cutri:2003}, 19-Derived from $L_{1\mu m}$ reported in \cite{Bosman:2023}, 20-\cite{Leipski:2014},
$\star$-\cite{Schlafly:2019}, $\ast$-\cite{Cutri:2013},
$\dag$-This work.\\}
\label{table:photometry_master}
\end{table}
\end{landscape}

\section{E-XQR-30 magnitudes and derived properties}
We collect the optical-to-MIR photometry taken from the literature for the QSOs in the extended XQR-30 sample.
Tab. \ref{table:photometry_xqr30} presents the data in band z, Y, J, H, K, W1, W2, W3 and W4.
If available,  the photometric data from Spitzer in the IRAC bands (3.6, 4.5 and 5.8 \textmu m) and  in the MIPS band at 24 \textmu m are also listed.
Tab. \ref{table:xqr30_luminosities} presents the derived bolometric and monochromatic luminosities for the sample while Fig. \ref{fig:xqr_30_sed} presents the results of the fits. In Tab. \ref{table:xqr30_spectral_slopes} we report the computed EUV, UV and optical spectral slopes.

\begin{landscape}
\begin{table}
\addtolength{\tabcolsep}{-2.5pt} 
\renewcommand\arraystretch{1.0}
\caption{Redshift and AB magnitudes of the E-XQR-30 QSOs.}
\begin{tabular}{lccccccccccc}

\toprule

\multirow{2}{*}{Name} & \multirow{2}{*}{Redshift$^{a}$} & \multirow{2}{*}{z$^{b}$} & \multirow{2}{*}{Y$^{b}$} & \multirow{2}{*}{J$^{c}$} & \multirow{2}{*}{H$^{c}$} & \multirow{2}{*}{K$^{c}$} & W1$^{b}$ & W2$^{b}$ & W3$^{b}$ & W4$^{b}$ & \multirow{2}{*}{S$_{24\mu m}$$^{d}$} \\
 & & & & & & & S$_{3.6\mu m}$$^{d}$ & S$_{4.5\mu m}$$^{d}$ & S$_{5.8\mu m}$$^{d}$ & S$_{8.0\mu m}$$^{d}$ & \\
\midrule
DELSJ0923+0402 & 6.633 &     & 21.03 $\pm$ 0.12 & 20.14 $\pm$ 0.08 &  19.9 $\pm$ 0.13 & 19.44 $\pm$ 0.07 & 19.23 $\pm$ 0.04 & 19.11 $\pm$ 0.07 &      &    &  \\
PSOJ323+12 &  6.587 &    &      & 19.74 $\pm$ 0.03 & 19.65 $\pm$ 0.06 & 19.21 $\pm$ 0.02 & 18.74 $\pm$ 0.02 & 18.91 $\pm$ 0.05 &      &     & \\
WISEAJ0439+1634 &   6.519 &   &      & 17.47 $\pm$ 0.02 & 17.33 $\pm$ 0.17 & 16.85 $\pm$ 0.01 & 16.68 $\pm$ 0.01 & 16.61 $\pm$ 0.01 & 15.45 $\pm$ 0.09 & 13.78 $\pm$ 0.12 & \\
PSOJ183+05 &  6.439 &    & 19.86 $\pm$ 0.09 & 19.77 $\pm$ 0.08 & 19.54 $\pm$ 0.11 & 19.54 $\pm$ 0.11 & 19.66 $\pm$ 0.05 & 19.76 $\pm$ 0.12 &      &      &\\
DELSJ1535+1943 &   6.37 &   &      & 19.63 $\pm$ 0.11 & 19.01 $\pm$ 0.05 &      & 18.58 $\pm$ 0.02 & 18.52 $\pm$ 0.03 & 17.76 $\pm$ 0.35 & 15.93 $\pm$ 0.45 &\\
VDESJ2211-3206 & 6.339 &     &      & 19.62 $\pm$ 0.03 & 19.43 $\pm$ 0.04 &  19.0 $\pm$ 0.03 &      &      &      &     & \\
  
\multirow{2}{*}{SDSSJ1030+0524} & \multirow{2}{*}{6.304} &     & \multirow{2}{*}{19.97 $\pm$ 0.06} & \multirow{2}{*}{19.87 $\pm$ 0.08} & \multirow{2}{*}{19.79 $\pm$ 0.11} & \multirow{2}{*}{19.49 $\pm$ 0.07} & 19.21 $\pm$ 0.04 & 19.07 $\pm$ 0.07 &      &     & \multirow{2}{*}{17.33 $\pm$ 0.15} \\
                & & & & & & & 19.23 $\pm$ 0.04 & 19.01 $\pm$ 0.02 & 19.61 $\pm$ 0.15 & 19.10 $\pm$ 0.12 &  \\
  
PSOJ065-26 &   6.187 &   &      & 19.36 $\pm$ 0.02 & 19.15 $\pm$ 0.04 & 19.22 $\pm$ 0.06 & 19.01 $\pm$ 0.03 &  19.2 $\pm$ 0.06 & 17.94 $\pm$ 0.53 &      &\\
PSOJ060+24 & 6.18 & 20.32 $\pm$ 0.09 & 20.28 $\pm$ 0.12 & 19.71 $\pm$ 0.05 & 19.72 $\pm$ 0.08 & 19.53 $\pm$ 0.06 & 19.21 $\pm$ 0.04 & 19.22 $\pm$ 0.08 &      &      &\\
PSOJ359-06 & 6.172 &     & 20.07 $\pm$ 0.12 &  19.85 $\pm$ 0.1 &  19.7 $\pm$ 0.02 & 19.32 $\pm$ 0.02 & 19.19 $\pm$ 0.03 &  19.0 $\pm$ 0.06 &      &      &\\
PSOJ217-07 &  6.166 &    & 20.33 $\pm$ 0.12 & 19.85 $\pm$ 0.08 & 19.99 $\pm$ 0.04 & 19.82 $\pm$ 0.15 & 19.95 $\pm$ 0.07 & 19.72 $\pm$ 0.11 &      &      &\\
PSOJ217-16 &  6.15 &    & 20.02 $\pm$ 0.08 & 19.71 $\pm$ 0.05 & 19.73 $\pm$ 0.11 & 19.37 $\pm$ 0.09 & 19.12 $\pm$ 0.03 &  19.1 $\pm$ 0.07 &      &     & \\
ULASJ1319+0950 &   6.135 &   & 19.84 $\pm$ 0.06 & 19.58 $\pm$ 0.04 & 19.71 $\pm$ 0.13 & 19.44 $\pm$ 0.13 &  19.6 $\pm$ 0.04 & 19.59 $\pm$ 0.09 &      &     & \\
CFHQSJ1509-1749 &   6.123 &   &  19.8 $\pm$ 0.04 & 19.75 $\pm$ 0.05 &      & 19.49 $\pm$ 0.09 &      & 19.31 $\pm$ 0.09 &      &      &\\
PSOJ239-07 &   6.110 &   & 19.56 $\pm$ 0.07 & 19.37 $\pm$ 0.07 & 19.27 $\pm$ 0.01 & 18.98 $\pm$ 0.09 & 18.97 $\pm$ 0.03 & 18.71 $\pm$ 0.05 &      &      &\\
  
\multirow{2}{*}{SDSSJ0842+1218} & \multirow{2}{*}{6.075} &     &      & \multirow{2}{*}{19.78 $\pm$ 0.03} &      & \multirow{2}{*}{19.22 $\pm$ 0.04} & 19.03 $\pm$ 0.03 & 19.09 $\pm$ 0.07 & 17.53 $\pm$ 0.43 &      & \multirow{2}{*}{16.12 $\pm$ 0.06}\\
& & & & & & & 19.13 $\pm$ 0.01 & 18.92 $\pm$ 0.02 & 19.03 $\pm$ 0.1 & 18.63 $\pm$ 0.08 &  \\

PSOJ158-14 &  6.069 &    & 19.28 $\pm$ 0.07 & 19.25 $\pm$ 0.08 &  19.15 $\pm$ 0.1 & 18.68 $\pm$ 0.09 & 18.65 $\pm$ 0.02 & 18.52 $\pm$ 0.04 &  16.6 $\pm$ 0.18 &     & \\
VDESJ0408-5632 & 6.035 &     &      & 19.85 $\pm$ 0.05 & 19.68 $\pm$ 0.05 & 19.63 $\pm$ 0.05 & 20.07 $\pm$ 0.04 & 19.91 $\pm$ 0.08 &      &      &\\

\multirow{2}{*}{SDSSJ1306+0356} & \multirow{2}{*}{6.033} & \multirow{2}{*}{19.53 $\pm$ 0.02} & \multirow{2}{*}{19.97 $\pm$ 0.03} & \multirow{2}{*}{19.69 $\pm$ 0.03} & \multirow{2}{*}{19.65 $\pm$ 0.05} & \multirow{2}{*}{19.34 $\pm$ 0.04} & 19.66 $\pm$ 0.05 &  19.57 $\pm$ 0.1 & 17.45 $\pm$ 0.38 &     & \multirow{2}{*}{17.49 $\pm$ 0.16}\\
& & & & & & & 19.24 $\pm$ 0.04 & 19.13 $\pm$ 0.04 & 19.57 $\pm$ 0.14 & 19.51 $\pm$ 0.13 &  \\
     
PSOJ009-10 &  6.004 &    &      & 19.93 $\pm$ 0.07 & 19.55 $\pm$ 0.05 & 19.24 $\pm$ 0.07 & 19.19 $\pm$ 0.03 & 19.06 $\pm$ 0.06 &      &      &\\
SDSSJ2310+1855 &  6.003 &    &      & 18.86 $\pm$ 0.05 & 18.94 $\pm$ 0.04 & 18.96 $\pm$ 0.05 & 18.57 $\pm$ 0.02 & 18.64 $\pm$ 0.04 & 17.45 $\pm$ 0.44 &     & \\
PSOJ007+04 &   6.002 &   & 20.22 $\pm$ 0.09 & 19.93 $\pm$ 0.09 & 20.05 $\pm$ 0.03 & 19.79 $\pm$ 0.05 & 19.88 $\pm$ 0.06 & 19.94 $\pm$ 0.13 &      &      &\\
PSOJ029-29 & 5.984 & 19.27 $\pm$ 0.01 & 19.21 $\pm$ 0.02 & 19.07 $\pm$ 0.04 & 19.14 $\pm$ 0.02 & 18.91 $\pm$ 0.03 & 18.95 $\pm$ 0.03 & 18.84 $\pm$ 0.05 & 17.18 $\pm$ 0.29 &      &\\
ULASJ0148+0600 & 5.977 & 19.23 $\pm$ 0.01 & 19.52 $\pm$ 0.06 & 19.11 $\pm$ 0.02 & 19.12 $\pm$ 0.07 & 19.02 $\pm$ 0.08 & 18.81 $\pm$ 0.02 &  18.7 $\pm$ 0.05 &  17.35 $\pm$ 0.3 & 15.58 $\pm$ 0.39 &\\
VDESJ2250-5015 & 5.977 &     &      & 19.17 $\pm$ 0.04 & 19.12 $\pm$ 0.05 & 18.94 $\pm$ 0.05 & 18.95 $\pm$ 0.02 & 18.91 $\pm$ 0.05 & 17.13 $\pm$ 0.26 &      &\\
  
\multirow{2}{*}{SDSSJ0818+1722} & \multirow{2}{*}{5.96}  &     &      & \multirow{2}{*}{19.09 $\pm$ 0.06} &      &      & 18.31 $\pm$ 0.02 & 18.21 $\pm$ 0.03 &  17.1 $\pm$ 0.33 &      &\multirow{2}{*}{16.40 $\pm$ 0.04}\\
 & & & & & & & 18.35 $\pm$ 0.01 & 18.14 $\pm$ 0.01 & 18.35 $\pm$ 0.05 & 18.08 $\pm$ 0.05 &  \\

PSOJ089-15 &  5.957 &    &      & 19.17 $\pm$ 0.04 & 18.78 $\pm$ 0.03 & 18.43 $\pm$ 0.04 & 18.21 $\pm$ 0.01 & 17.86 $\pm$ 0.02 & 17.24 $\pm$ 0.28 &      &\\
PSOJ108+08 &   5.949 &   &      & 19.07 $\pm$ 0.06 & 19.03 $\pm$ 0.07 &      & 18.72 $\pm$ 0.02 & 18.46 $\pm$ 0.04 &  17.7 $\pm$ 0.54 &      &\\
PSOJ023-02 & 5.9 &     &  20.2 $\pm$ 0.16 & 20.06 $\pm$ 0.06 & 20.08 $\pm$ 0.04 &  19.7 $\pm$ 0.17 & 19.27 $\pm$ 0.03 & 18.78 $\pm$ 0.05 &      &      &\\
PSOJ183-12 &  5.86 &    & 19.23 $\pm$ 0.04 & 19.07 $\pm$ 0.03 &  19.1 $\pm$ 0.04 &  19.1 $\pm$ 0.08 & 18.96 $\pm$ 0.03 &  19.2 $\pm$ 0.08 &      &      &\\
PSOJ025-11 &  5.85 &    &      & 19.65 $\pm$ 0.06 & 19.78 $\pm$ 0.07 &  19.6 $\pm$ 0.04 & 19.46 $\pm$ 0.04 & 19.24 $\pm$ 0.07 & 17.43 $\pm$ 0.43 &      &\\
PSOJ242-12 &   5.83  &    &      & 19.71 $\pm$ 0.09 & 19.66 $\pm$ 0.02 & 19.46 $\pm$ 0.26 & 19.27 $\pm$ 0.04 &  19.56 $\pm$ 0.1 & 17.04 $\pm$ 0.32 &     &\\
PSOJ308-27 &  5.80 &    &  19.64 $\pm$ 0.1 &  19.5 $\pm$ 0.04 & 19.62 $\pm$ 0.05 & 19.63 $\pm$ 0.15 & 19.53 $\pm$ 0.05 & 19.22 $\pm$ 0.08 &      &     & \\
PSOJ065+01 & 5.80 &     &      & 19.74 $\pm$ 0.05 & 19.48 $\pm$ 0.03 & 19.37 $\pm$ 0.05 &      &      &      &      &\\
  
\multirow{2}{*}{SDSSJ0836+0054} & \multirow{2}{*}{5.773} &  \multirow{2}{*}{18.81 $\pm$ 0.01} &  \multirow{2}{*}{18.88 $\pm$ 0.01} &  \multirow{2}{*}{18.65 $\pm$ 0.01} &  \multirow{2}{*}{18.88 $\pm$ 0.01} & \multirow{2}{*}{18.36 $\pm$ 0.01} & 17.97 $\pm$ 0.01 & 17.77 $\pm$ 0.02 & 17.74 $\pm$ 0.53 &      &\multirow{2}{*}{16.48 $\pm$ 0.06}\\
& & & & & & & 17.87 $\pm$ 0.01 & 17.49 $\pm$ 0.01$^{d}$ & 17.77 $\pm$ 0.02 & 17.70 $\pm$ 0.02 &  \\
  
\multirow{2}{*}{SDSSJ0927+2001} & \multirow{2}{*}{5.772} &    &      &  \multirow{2}{*}{19.95 $\pm$ 0.1} &      &      & 19.61 $\pm$ 0.05 & 19.92 $\pm$ 0.14 &      &      & \multirow{2}{*}{16.88$\pm$0.08}\\
& & & & & & & 19.72 $\pm$ 0.05 & 19.65 $\pm$ 0.04 & 19.82 $\pm$ 0.18 & 19.22 $\pm$ 0.1 &  \\
\bottomrule
\end{tabular}
\addtolength{\tabcolsep}{-2.5pt} 
\tablefoot{References: $^{a}$-\cite{Mazzucchelli:2023}, $^{b}$-\cite{Ross:2015}, $^{b}$-\cite{DOdorico:2023},  $^{c}$-\cite{Leipski:2014}, $^{d}$-\cite{Jiang:2006}.}
\label{table:photometry_xqr30}
\end{table}
\end{landscape}

\begin{table*}[h]
\addtolength{\tabcolsep}{-1.1pt} 
\begin{tabular}{lcccccccc}
\toprule
              Name &             $L_\text{bol}$ &             $\lambda L_{2500\text{\AA}}$ &            $\lambda L_{3000\text{\AA}}$ &            $\lambda L_{4400\text{\AA}}$ &            $\lambda L_{5100\text{\AA}}$ &  E[B-V] &
              $\lambda_\text{Edd}$ & $\lambda L_{1keV}$ \\
\midrule
DELSJ0923+0402	&	47.33 $\pm $ 0.05	&	46.90 $\pm $ 0.02	&	46.86 $\pm $ 0.02	&	46.69 $\pm $ 0.01	&	46.63 $\pm $ 0.02	&	0.09	&	0.64 $\pm $ 0.30	&	45.22 \\
PSOJ323+12	&	47.37 $\pm $ 0.05	&	46.90 $\pm $ 0.01	&	46.88 $\pm $ 0.01	&	46.83 $\pm $ 0.01	&	46.73 $\pm $ 0.01	&	0.06	&	2.22 $\pm $ 0.60	&	45.30 \\
WISEAJ0439+1634$^{\dag}$	&	48.33 $\pm $ 0.05	&	47.89 $\pm $ 0.02	&	47.86 $\pm $ 0.01	&	47.68 $\pm $ 0.01	&	47.61 $\pm $ 0.01	&	0.08	&	3.23 $\pm $ 0.71	&	45.97 \\
PSOJ183+05	&	47.11 $\pm $ 0.04	&	46.67 $\pm $ 0.04	&	46.59 $\pm $ 0.04	&	46.37 $\pm $ 0.02	&	46.29 $\pm $ 0.03	&	0.01	&	0.40 $\pm $ 0.29	&	45.19 \\
DELSJ1535+1943	&	47.50 $\pm $ 0.07	&	47.07 $\pm $ 0.01	&	47.00 $\pm $ 0.01	&	46.87 $\pm $ 0.01	&	46.81 $\pm $ 0.01	&	0.06	&	0.40 $\pm $ 0.08	&	45.52 \\
VDESJ2211-3206	&	47.34 $\pm $ 0.08	&	46.88 $\pm $ 0.01	&	46.80 $\pm $ 0.08$^{a}$	&	46.61 $\pm $ 0.08$^{a}$	&	46.56 $\pm $ 0.08$^{a}$	&	0.04	&	0.86 $\pm $ 0.42	&	45.28 \\
SDSSJ1030+0524	&	47.17 $\pm $ 0.04	&	46.66 $\pm $ 0.03	&	46.64 $\pm $ 0.02	&	46.54 $\pm $ 0.02	&	46.51 $\pm $ 0.01	&	0.03	&	0.63 $\pm $ 0.14	&	45.19 \\
PSOJ065-26	&	47.26 $\pm $ 0.04	&	46.80 $\pm $ 0.02	&	46.70 $\pm $ 0.02	&	46.60 $\pm $ 0.01	&	46.52 $\pm $ 0.01	&	0.01	&	0.40 $\pm $ 0.15	&	45.35 \\
PSOJ060+24	&	47.21 $\pm $ 0.05	&	46.73 $\pm $ 0.02	&	46.67 $\pm $ 0.02	&	46.57 $\pm $ 0.01	&	46.51 $\pm $ 0.02	&	0.05	&	0.75 $\pm $ 0.16	&	45.11 \\
PSOJ359-06	&	47.22 $\pm $ 0.05	&	46.74 $\pm $ 0.01	&	46.73 $\pm $ 0.01	&	46.58 $\pm $ 0.01	&	46.53 $\pm $ 0.01	&	0.04	&	1.31 $\pm $ 0.36	&	45.19 \\
PSOJ217-07	&	47.00 $\pm $ 0.05	&	46.50 $\pm $ 0.03	&	46.46 $\pm $ 0.05	&	46.24 $\pm $ 0.03	&	46.19 $\pm $ 0.03	&	0.01	&	1.00 $\pm $ 0.50	&	45.06 \\
PSOJ217-16	&	47.19 $\pm $ 0.06	&	46.69 $\pm $ 0.03	&	46.67 $\pm $ 0.03	&	46.58 $\pm $ 0.02	&	46.52 $\pm $ 0.02	&	0.03	&	1.18 $\pm $ 0.74	&	45.21 \\
ULASJ1319+0950	&	47.12 $\pm $ 0.04	&	46.61 $\pm $ 0.05	&	46.59 $\pm $ 0.05	&	46.37 $\pm $ 0.02	&	46.30 $\pm $ 0.02	&	0.01	&	0.31 $\pm $ 0.06	&	45.25 \\
CFHQSJ1509-1749	&	47.12 $\pm $ 0.05	&	46.63 $\pm $ 0.03	&	46.58 $\pm $ 0.03	&	46.45 $\pm $ 0.04	&	46.39 $\pm $ 0.04	&	0.01	&	0.53 $\pm $ 0.20	&	45.22 \\
PSOJ239-07	&	47.30 $\pm $ 0.05	&	46.83 $\pm $ 0.01	&	46.81 $\pm $ 0.03	&	46.64 $\pm $ 0.01	&	46.59 $\pm $ 0.01	&	0.02	&	0.60 $\pm $ 0.10	&	45.33 \\
SDSSJ0842+1218	&	47.39 $\pm $ 0.04	&	46.92 $\pm $ 0.01	&	46.86 $\pm $ 0.01	&	46.66 $\pm $ 0.01	&	46.59 $\pm $ 0.01	&	0.09	&	0.98 $\pm $ 0.20	&	45.22 \\
PSOJ158-14	&	47.45 $\pm $ 0.06	&	46.95 $\pm $ 0.03	&	46.96 $\pm $ 0.03	&	46.79 $\pm $ 0.01	&	46.73 $\pm $ 0.01	&	0.04	&	1.10 $\pm $ 0.25	&	45.35 \\
VDESJ0408-5632	&	47.03 $\pm $ 0.03	&	46.58 $\pm $ 0.02	&	46.51 $\pm $ 0.02	&	46.20 $\pm $ 0.02	&	46.11 $\pm $ 0.02	&	0.01	&	0.42 $\pm $ 0.09	&	45.09 \\
SDSSJ1306+0356	&	47.11 $\pm $ 0.04	&	46.63 $\pm $ 0.02	&	46.62 $\pm $ 0.02	&	46.49 $\pm $ 0.02	&	46.44 $\pm $ 0.02	&	0.01	&	0.53 $\pm $ 0.17	&	45.21 \\
PSOJ009-10	&	47.25 $\pm $ 0.06	&	46.81 $\pm $ 0.01	&	46.77 $\pm $ 0.02	&	46.59 $\pm $ 0.01	&	46.52 $\pm $ 0.01	&	0.06	&	0.18 $\pm $ 0.04	&	45.54 \\
SDSSJ2310+1855	&	47.37 $\pm $ 0.04	&	46.87 $\pm $ 0.02	&	46.79 $\pm $ 0.02	&	46.73 $\pm $ 0.01	&	46.69 $\pm $ 0.01	&	0.01	&	0.40 $\pm $ 0.07	&	45.45 \\
PSOJ007+04	&	46.96 $\pm $ 0.04	&	46.46 $\pm $ 0.01	&	46.44 $\pm $ 0.02	&	46.25 $\pm $ 0.02	&	46.17 $\pm $ 0.03	&	0.01	&	0.09 $\pm $ 0.02	&	45.54 \\
PSOJ029-29	&	47.33 $\pm $ 0.05	&	46.84 $\pm $ 0.01	&	46.81 $\pm $ 0.01	&	46.63 $\pm $ 0.01	&	46.57 $\pm $ 0.01	&	0.02	&	0.74 $\pm $ 0.15	&	45.35 \\
ULASJ0148+0600	&	47.37 $\pm $ 0.06	&	46.88 $\pm $ 0.02	&	46.80 $\pm $ 0.03	&	46.68 $\pm $ 0.01	&	46.64 $\pm $ 0.01	&	0.03	&	0.45 $\pm $ 0.09	&	45.39 \\
VDESJ2250-5015	&	47.31 $\pm $ 0.05	&	46.84 $\pm $ 0.02	&	46.79 $\pm $ 0.02	&	46.62 $\pm $ 0.01	&	46.56 $\pm $ 0.01	&	0.02	&	0.36 $\pm $ 0.34	&	45.39 \\
SDSSJ0818+1722	&	47.53 $\pm $ 0.03	&	47.05 $\pm $ 0.01	&	46.99 $\pm $ 0.01	&	46.88 $\pm $ 0.01	&	46.85 $\pm $ 0.01	&	0.06	&	0.47 $\pm $ 0.08	&	45.51 \\
PSOJ089-15	&	47.64 $\pm $ 0.04	&	47.20 $\pm $ 0.01	&	47.15 $\pm $ 0.01	&	47.00 $\pm $ 0.01	&	46.96 $\pm $ 0.01	&	0.09	&	0.94 $\pm $ 0.22	&	45.46 \\
PSOJ108+08	&	47.38 $\pm $ 0.06	&	46.89 $\pm $ 0.02	&	46.83 $\pm $ 0.02	&	46.71 $\pm $ 0.01	&	46.67 $\pm $ 0.01	&	0.02	&	0.61 $\pm $ 0.15	&	45.39 \\
PSOJ023-02	&	47.21 $\pm $ 0.06	&	46.67 $\pm $ 0.02	&	46.63 $\pm $ 0.04	&	46.55 $\pm $ 0.02	&	46.54 $\pm $ 0.01	&	0.09	&	0.53 $\pm $ 0.10	&	45.14 \\
PSOJ183-12	&	47.27 $\pm $ 0.04	&	46.79 $\pm $ 0.02	&	46.71 $\pm $ 0.03	&	46.58 $\pm $ 0.02	&	46.51 $\pm $ 0.01	&	0.01	&	0.88 $\pm $ 0.30	&	45.31 \\
PSOJ025-11	&	47.08 $\pm $ 0.05	&	46.56 $\pm $ 0.02	&	46.51 $\pm $ 0.02	&	46.39 $\pm $ 0.02	&	46.35 $\pm $ 0.02	&	0.02	&	0.43 $\pm $ 0.08	&	45.17 \\
PSOJ242-12	&	47.14 $\pm $ 0.08	&	46.65 $\pm $ 0.02	&	46.60 $\pm $ 0.07	&	46.48 $\pm $ 0.04	&	46.42 $\pm $ 0.02	&	0.03	&	0.34 $\pm $ 0.11	&	45.22 \\
PSOJ308-27	&	47.09 $\pm $ 0.04	&	46.58 $\pm $ 0.03	&	46.49 $\pm $ 0.05	&	46.35 $\pm $ 0.03	&	46.31 $\pm $ 0.02	&	0.01	&	0.79 $\pm $ 0.11	&	45.16 \\
PSOJ065+01	&	47.23 $\pm $ 0.09	&	46.77 $\pm $ 0.01	&	46.69 $\pm $ 0.01	&	46.53 $\pm $ 0.02	&	46.47 $\pm $ 0.03	&	0.05	&	0.33 $\pm $ 0.13	&	45.31 \\
SDSSJ0836+0054	&	47.58 $\pm $ 0.03	&	47.01 $\pm $ 0.01	&	47.03 $\pm $ 0.01	&	47.01 $\pm $ 0.01	&	46.99 $\pm $ 0.01	&	0.05	&	0.77 $\pm $ 0.19	&	45.48 \\
SDSSJ0927+2001	&	47.06 $\pm $ 0.04	&	46.57 $\pm $ 0.03	&	46.48 $\pm $ 0.02	&	46.31 $\pm $ 0.02	&	46.25 $\pm $ 0.02	&	0.04	&	0.71 $\pm $ 0.17	&	45.08 \\

\bottomrule
\end{tabular}
\caption{Same as in Tab. \ref{table:luminosities} for the E-XQR-30 sample. The Eddington ratios have been computed from the MgII estimated BH masses reported in \cite{Mazzucchelli:2023}. $\lambda L_{1keV}$ has been estimated with SED fitting with QSOSED models as described in Sec. \ref{sec:qsosed}.\\
$^{\dag}$ This QSO is known to be gravitationally lensed \citep[][]{Fan:2019} and therefore its luminosities are magnified.\\
$^{a}$	For this QSO there are not sufficient photometric points to derive monochromatic luminosities in these wavelengths. The values reported have been derived interpolating the best-fit SED.}
\label{table:xqr30_luminosities}
\end{table*}

\begin{table}
\centering
\begin{tabular}{lccc}
\toprule
Name & $\alpha_{EUV}$	 & $\beta_{UV}$ & $\gamma_{opt}$ \\
\midrule
DELSJ0923+0402	&	-2.02	&	-1.58 $\pm $ 0.16	&	 \\
PSOJ323+12	&	-2.04	&	-0.87 $\pm $ 0.13	&	 \\
WISEAJ0439+1634	&	-2.66	&	-1.00 $\pm $ 0.12	&	 \\
PSOJ183+05	&	-1.88	&	-0.43 $\pm $ 0.18	&	 \\
DELSJ1535+1943	&	-2.13	&	 	&	             \\
VDESJ2211-3206	&	-2.02	&	-1.01 $\pm $ 0.13	&	 \\ 
SDSSJ1030+0524	&	-1.91	&	-0.37 $\pm $ 0.28	&	-0.31 $\pm $ 0.10 \\
PSOJ065-26	&	-1.97	&	 	&	-0.07 $\pm $ 0.14                      \\
PSOJ060+24	&	-1.94	&	-0.77 $\pm $ 0.30	&	-0.44 $\pm $ 0.15       \\
PSOJ359-06	&	-1.95	&	-0.67 $\pm $ 0.25	&	-0.37 $\pm $ 0.11 \\
PSOJ217-07	&	-1.8	&	-0.34 $\pm $ 0.25	&	-0.15 $\pm $ 0.25  \\
PSOJ217-16	&	-1.93	&	-0.62 $\pm $ 0.30	&	-0.32 $\pm $ 0.17   \\
ULASJ1319+0950	&	-1.88	&	-0.47 $\pm $ 0.30	&	0.17 $\pm $ 0.21 \\
CFHQSJ1509-1749	&	-1.88	&	 	&	  \\
PSOJ239-07	&	-2.0	&	-0.53 $\pm $ 0.19	&	-0.38 $\pm $ 0.15 \\
SDSSJ0842+1218	&	-2.06	&	 	&	-0.39 $\pm $ 0.08 \\
PSOJ158-14	&	-2.1	&	-0.26 $\pm $ 0.28	&	-0.22 $\pm $ 0.15\\
VDESJ0408-5632	&	-1.82	&	 	&	0.46 $\pm $ 0.14 \\
SDSSJ1306+0356	&	-1.88	&	-0.64 $\pm $ 0.18	&	-0.00 $\pm $ 0.09 \\
PSOJ009-10	&	-1.97	&	 	&	-0.22 $\pm $ 0.15 \\
SDSSJ2310+1855	&	-2.05	&	 	&	-0.41 $\pm $ 0.12 \\
PSOJ007+04	&	-1.78	&	-0.20 $\pm $ 0.22	&	0.19 $\pm $ 0.17 \\
PSOJ029-29	&	-2.02	&	-0.12 $\pm $ 0.15	&	-0.07 $\pm $ 0.11 \\
ULASJ0148+0600	&	-2.04	&	-0.75 $\pm $ 0.23	&	-0.40 $\pm $ 0.14 \\
VDESJ2250-5015	&	-2.01	&	 	&	-0.04 $\pm $ 0.12 \\
SDSSJ0818+1722	&	-2.15	&	 	&	-0.23 $\pm $ 0.10 \\
PSOJ089-15	&	-2.22	&	 	&	-0.71 $\pm $ 0.10 \\
PSOJ108+08	&	-2.05	&	 	&	  \\
PSOJ023-02	&	-1.94	&	-0.08 $\pm $ 0.28	&	-1.25 $\pm $ 0.20 \\
PSOJ183-12	&	-1.98	&	-0.24 $\pm $ 0.18	&	0.10 $\pm $ 0.16 \\
PSOJ025-11	&	-1.85	&	 	&	-0.42 $\pm $ 0.13 \\
PSOJ242-12	&	-1.89	&	 	&	0.44 $\pm $ 0.30 \\
PSOJ308-27	&	-1.86	&	0.11 $\pm $ 0.24	&	-0.59 $\pm $ 0.22 \\
PSOJ065+01	&	-1.95	&	 	&	  \\
SDSSJ0836+0054	&	-2.18	&	0.01 $\pm $ 0.10	&	-0.71 $\pm $ 0.07 \\
SDSSJ0927+2001	&	-1.85	&	 	&	-0.05 $\pm $ 0.30 \\

\bottomrule
\end{tabular}
\caption{Same as in Tab. \ref{table:spectral_slopes} for the E-XQR-30 sources. $\alpha_{EUV}$ has been computed assuming the mean $\lambda L_{1keV}$ of the HYPERION QSOs.}
\label{table:xqr30_spectral_slopes}
\end{table}

\begin{figure*}
    \centering
    \includegraphics[width=0.9\textwidth]{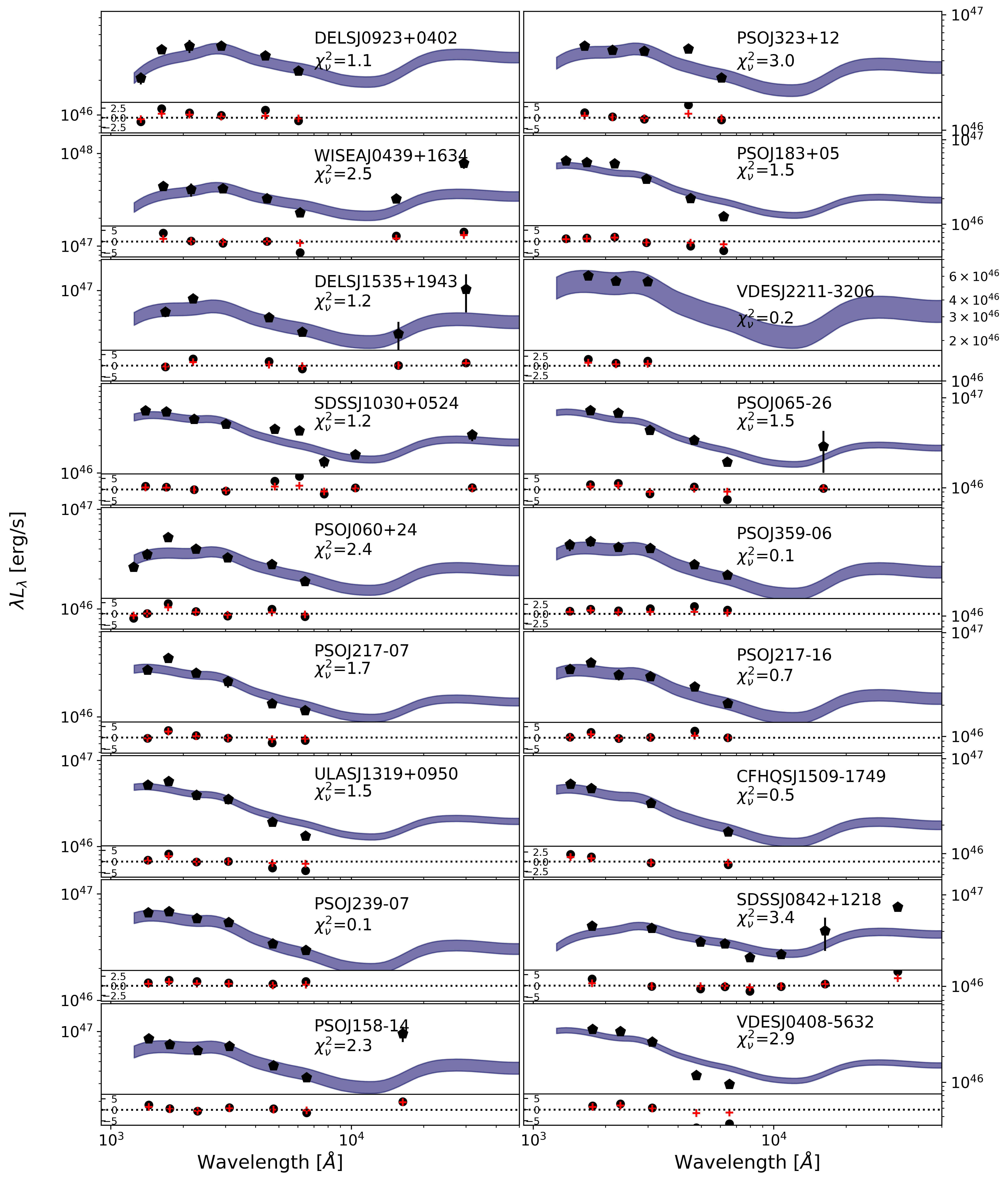}
        \caption{Results of the SED fitting of the E-XQR-30 QSOs using the lum-K13 template. Lower sub-panels show the residuals, assuming only the uncertainties on the data points (black circles) or taking also into account the scatter in the SED (red crosses).}
        \label{fig:xqr_30_sed}
\end{figure*}

\begin{figure*} \ContinuedFloat
    \centering
    \includegraphics[width=0.9\textwidth]{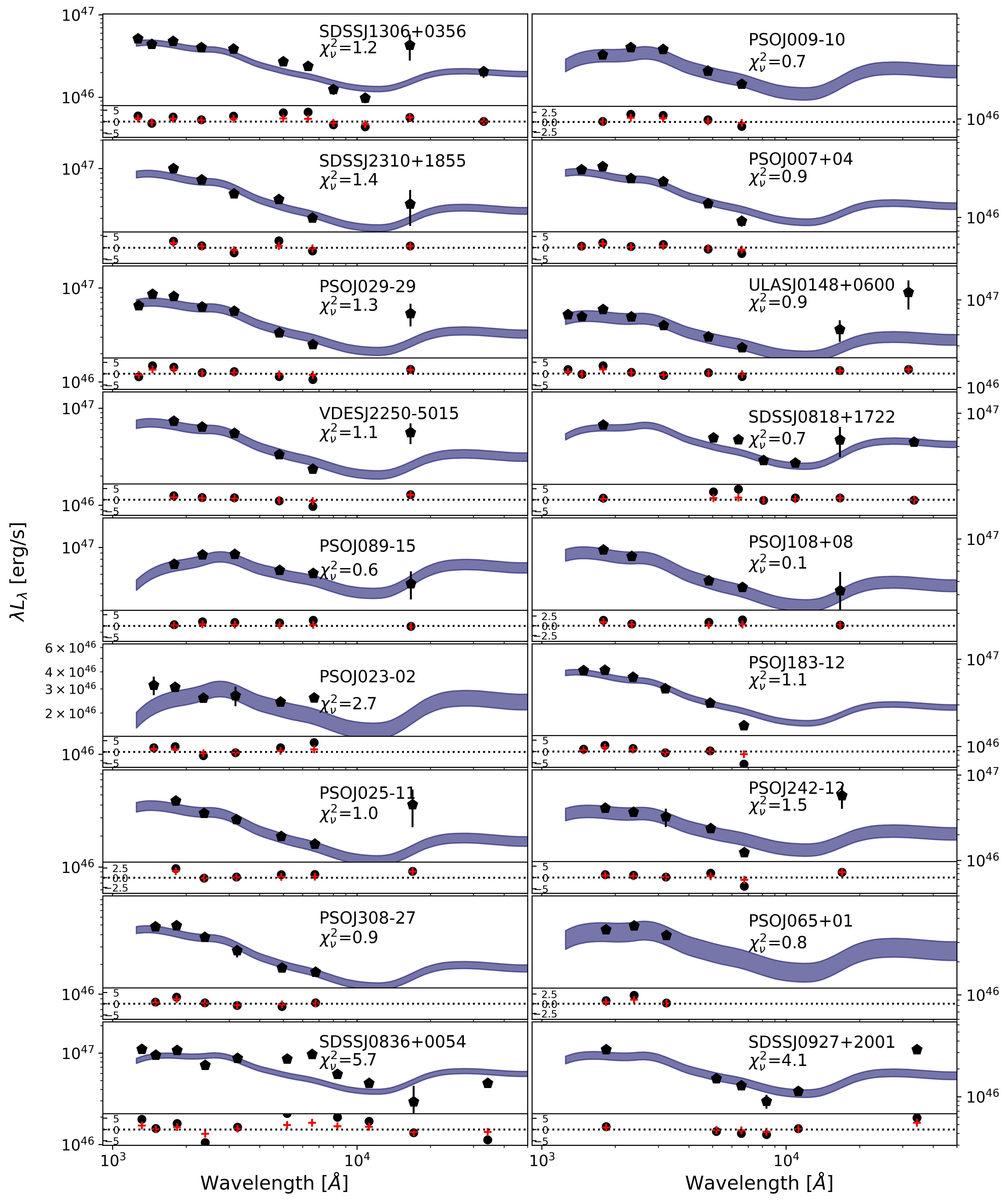}
    \caption{Continued}
    \label{fig:xqr_30_sed_pt2}
\end{figure*}

\end{document}